\documentstyle[preprint,eqsecnum,aps]{revtex}
\newif\iftightenlines\tightenlinesfalse
\tightenlines\tightenlinestrue
\begin{document}
%
%%%%%%%%%%%%%%%%%%%%%%%%%%%%%%%%%%%%%%%%%%%%%%%%%%%%%%%%%%%%%%%%%
%
%%%%%%%%%%%%%%%%%%%% TITLE PAGE %%%%%%%%%%%%%%%%%%%%%%%%%%%%%%%%%
%
\draft
\preprint{
   \vbox{\baselineskip=14pt%
   \rightline{ITP-SB-02-98}\break
   \rightline{INLO-PUB-02/98} }
}

\title{
TWO-LOOP OPERATOR MATRIX ELEMENTS CALCULATED UP TO FINITE TERMS
FOR POLARIZED DEEP INELASTIC LEPTON-HADRON SCATTERING}

\author{
Y. Matiounine$^1$, J. Smith$^1$ and W.L. van Neerven$^2$
}

\address{
$^1$Institute for Theoretical Physics,
State University of New York at Stony Brook,
Stony Brook, NY 11794-3840, USA
}
\address{
$^2$Instituut-Lorentz,
University of Leiden,
PO Box 9506, 2300 RA Leiden, The Netherlands
}
\date{March 1998}
\maketitle
\begin{abstract}
We present the two-loop corrected operator matrix elements
contributing to the scale evolution of the
longitudinal spin structure function $g_1(x,Q^2)$ calculated up 
to finite terms 
which survive in the limit $\varepsilon = N - 4 \rightarrow 0 $.
These terms are needed to renormalize the local operators
up to third order in the strong coupling constant $\alpha_s$. Further the  
expressions for the two-loop corrected operator matrix elements can be 
inserted into one loop graphs to obtain a part of the third order 
contributions to these matrix elements. 
This work is a first step in obtaining the third order anomalous dimensions
so that a complete next-to-next-to-leading order (NNLO) analysis of 
the above mentioned structure function can be carried out. 
In our calculation particular 
attention is paid to the renormalization constant which is needed to
restore the Ward-identities violated by the HVBM prescription for the 
$\gamma_5$-matrix in $N$-dimensional regularization.
\end{abstract}
\medskip
%%%%%%%%%%%%%%%%%%%%%%%%%%%%%%%%%%%%%%%%%%%%%%%%%%%%%%%
%% MAIN TEXT %%
%------------------This is Section 1---------------------------------
\section{Introduction}
%----------------------------------------------------------
During the last ten years there has been a lot of activity in the area of
spin physics in particular the study of the structure functions $g_i(x,Q^2)$
($i=1,2$) measured in polarized lepton-hadron scattering (for reviews 
see \cite{hera} and \cite{zeuth}). One of the 
achievements was the determination of the anomalous dimensions of 
twist two local composite operators up to two-loop 
order (see \cite{mene}, \cite{vogel}) which determine the scale ($Q^2$)
evolution of the longitudinal spin structure function $g_1(x,Q^2)$. 
Since also the order $\alpha_s$ contributions to the coefficient functions
are known (see \cite{koma}, \cite{boqi}) it is now possible to make a
full next-to-leading order (NLO) analysis of $g_1(x,Q^2)$ analogous to what
has been done for the spin averaged structure function $F_2(x,Q^2)$. When the 
statistics of the ongoing experiments improve it will be also necessary
to investigate how the NLO predictions are modified by including yet
higher order corrections. In particular this will be interesting for the
study of the small $x$-region where gluon contributions to the anomalous 
dimensions and the coefficient functions are important. A start of this
programme, which will lead to a complete next-to-next-to-leading order (NNLO)
description of the longitudinal spin structure function, has been made in
\cite{zine} where quark as well as gluon coefficient functions have been
computed up to order $\alpha_s^2$. The three-loop
anomalous dimensions are still missing.

In this paper we want to make the first step in a three loop programme by 
computing the operator matrix elements (OME's), 
obtained by sandwiching the local composite 
operators between quark and gluon states, up to two-loop order including
non-pole terms which are finite in the limit $N \rightarrow 4$. Here $N$
refers to the method of $N$-dimensional regularization which is used to 
regularize the ultraviolet divergences occurring in the OME's. 
Using this method the latter divergences manifest themselves
as pole terms of the type $1/(N-4)^k$. These finite terms are needed to
renormalize the three-loop graphs contributing to the OME's. Further one 
can use the unrenormalized two-loop expressions to determine a part of
the three-loop contributions to the OME's by inserting them into one loop 
diagrams. Another interesting feature of the calculation of the spin OME's
is the appearance of the $\gamma_5$-matrix and the Levi-Civita tensor in
the operator vertices. Since we use $N$-dimensional regularization one has to
find a suitable prescription to extend these objects to $N$ dimensions. In 
this work we have chosen the HVBM prescription given first by 't Hooft and
Veltman \cite{hove} and worked out in more detail by Breitenlohner and 
Maison \cite{brma}. One of the characteristics of this 
prescription is that the $\gamma_5$-matrix commutes with the other 
$\gamma_{\mu}$ matrices for $N > 4$.  
Although this is a consistent 
scheme, since it preserves the cyclicity of the traces, it violates some 
Ward-identities which would be preserved with an 
anti-commuting $\gamma_5$-matrix. To restore these 
Ward-identities one has to introduce additional renormalization 
factors called $Z_{qq}^{5,r}$. These factors will be calculated up to 
order $\alpha_s^2$ for the non-singlet ($r=$ NS) and singlet ($r=$ S)
operators for arbitrary spin. 

The paper is organized as follows. In section 2 we present the
algebraic formulae for the unrenormalized OME's expressed in  
renormalization group coefficients. We determine the 
factors $Z^{5,r}_{qq}$ for $r=$ NS (nonsinglet) and S (singlet)
and compare our results with earlier calculations in the literature.
The long expressions for the spin OME's, which are the results of our
calculations, are given in Appendix A.
%
%------------------This is Section 1---------------------------------
\section{The calculation of the two-loop operator matrix elements}
%----------------------------------------------------------
In this section we will give an outline of the calculation of the 
OME's up to two-loop order.
The operators, which appear in polarized lepton-hadron scattering, can be
split into singlet and non-singlet parts 
with respect to the flavour group. In leading twist (namely two) 
the non-singlet quark operator of spin $n$ is given by
\begin{eqnarray}
\label{eqn:2.1}
O_{q,k}^{5,\mu_1,\mu_2 \cdots \mu_n}=\frac{1}{2} i^{n-1} {\cal S} \Big [ \bar 
\psi(x) \gamma_5 \gamma^{\mu_1} D^{\mu_2}
\cdots D^{\mu_n} \frac{\lambda_k}{2} \psi (x) + \mbox{trace terms} \Big ]\,.
\end{eqnarray}
In the singlet case there are two operators. The quark operator is 
represented by
\begin{eqnarray}
\label{eqn:2.2}
O_q^{5,\mu_1,\mu_2 \cdots \mu_n}=\frac{1}{2} i^{n-1} {\cal S} \Big [ \bar 
\psi(x) \gamma_5 \gamma^{\mu_1} D^{\mu_2}
\cdots D^{\mu_n} \psi (x) + \mbox{trace terms} \Big ] \,,
\end{eqnarray}
and the gluon operator is given by
\begin{eqnarray}
\label{eqn:2.3}
O_g^{5,\mu_1,\mu_2 \cdots \mu_n}=\frac{1}{2} i^{n-2} {\cal S} \Big [ 
\epsilon^{\mu_1\alpha\beta\gamma} F_{\beta \gamma}(x) D^{a, \mu_2}
\cdots D^{\mu_{n-1}} F_{\alpha}^{a,\mu_n}(x)   + \mbox{trace terms} 
\Big ]\,.
\end{eqnarray}
In the composite operators above $\psi$ and $F_{\mu\nu}^a$ stand for the
quark field and the gluon field tensor respectively.
The $\lambda_k$ in Eq. (\ref{eqn:2.1}) represent the generators of the flavour
group and the index $a$ in Eq. (\ref{eqn:2.3}) stands for the colour.
Further the above operators are irreducible tensors
with respect to the Lorentz group so that they have to be
symmetric and traceless in all their Lorentz indices $\mu_i$.
From the operators above one can derive the Feynman rules for the operator
vertices in the standard way (see e.g. \cite{mene} and  \cite{sasa}). This
derivation is facilitated if the operators are multiplied by the source
\begin{eqnarray}
\label{eqn:2.4}
J_{\mu_1\mu_2 \cdots \mu_n} = \Delta_{\mu_1}\Delta_{\mu_2}\cdots 
\Delta_{\mu_n} \,,
\end{eqnarray}
with $\Delta^2=0$ in order to eliminate the trace terms in 
Eqs. (\ref{eqn:2.1})-(\ref{eqn:2.3}). Hence all operator vertices 
in momentum space are multiplied by a factor $(\Delta\cdot p)^n$. For 
the computation of the OME's denoted by
\begin{eqnarray}
\label{eqn:2.5}
A_{ij}^5 = \langle j(p) \mid O_i^5 \mid j(p) \rangle
\end{eqnarray}
with $i,j = q,g$ we choose the Feynman gauge except for the one-loop graphs
for which we take the general covariant gauge. For this choice the gluon
propagator equals
\begin{eqnarray}
\label{eqn:2.6}
\Delta_{ab}^{\mu\nu}(k)= \frac{i \delta_{ab}}{k^2} \Big ( - g^{\mu\nu} +
(1- \xi ) \frac{k^{\mu}k^{\nu}}{k^2} \Big )\,.
\end{eqnarray}
The matrix element (\ref{eqn:2.5}) has to be considered as a connected Green
function with the external legs amputated but with the external self
energies of the partons $j$ included.
In this paper all quarks and gluons are taken to be massless
and the external momentum $p$ is off-shell ($p^2 < 0$) 
in order to get finite expressions for the OME's.
This choice implies that the OME's are not gauge invariant so that they
cease to be ordinary S-matrix elements. Therefore they acquire
unphysical parts which usually vanish by virtue of the 
equations of motion (EOM) if the external
legs are on-shell. Contrary to the spin averaged operators
treated e.g. in \cite{msn} there is no mixing between 
gauge invariant (GI) or physical (PHYS) operators and non gauge 
invariant (NGI) operators  (see \cite{brs}-\cite{hasm}).
Therefore the renormalization of the operators 
in Eqs. (\ref{eqn:2.1})-(\ref{eqn:2.3})  
is much easier than is the case for the spin averaged 
operators in \cite{msn}. On the other hand, due to the presence of the 
$\gamma_5$-matrix 
and the Levi-Civita tensors in the composite operators above, one has to
correct for spurious terms in the OME's if one adopts $N$-dimensional
regularization. 

The calculation of the Feynman graphs corresponding to the physical operators
in Eqs. (\ref{eqn:2.1})-(\ref{eqn:2.3}), which are depicted in 
the figures in \cite{mene}, 
proceeds in the standard way. The corresponding Feynman integrals reveal
ultraviolet divergences which are regularized using the method of 
$N$-dimensional regularization. In this way the above divergences show up
in the form of pole terms of the type $(1/\varepsilon)^k$ with $\varepsilon=
N-4$. In \cite{mene}, \cite{vogel} it was sufficient to evaluate the one-loop
graphs up to finite and the two-loop graphs up to single pole terms in order
to get the second order anomalous dimensions. Here we have to include terms
proportional to $\varepsilon$ in the one-loop expressions and the two-loop
graphs have to be computed up to finite terms. The way to compute the two-loop
Feynman integrals up to finite terms is presented in \cite{msn}, \cite{hane}, 
and Appendix B of \cite{bmsmn}. We used the program FORM \cite{jos}
to do the necessary algebra.

As has been mentioned above, the presence of the 
$\gamma_5$-matrix, which is essentially a four dimensional object,
will cause some technical problems
when one chooses the method of $N$-dimensional regularization.  
One has to find a prescription to give a suitable definition 
valid for all space-time dimensions. 
We will adopt the HVBM prescription given by 't Hooft and Veltman \cite{hove}
which has been worked out in more detail by Breitenlohner and Maison 
\cite{brma}. In order to facilitate the calculation of the OME's it is more 
convenient (see \cite{akde}, \cite{larin}) to replace the term 
$
{\Delta \hspace{-0.52em}/\hspace{0.1em}}
%\Ds 
\gamma_5$ appearing in the operator vertices 
(see Appendix A in \cite{mene}) by
\begin{eqnarray}
\label{eqn:2.7}
{\Delta \hspace{-0.52em}/\hspace{0.1em}}
%\Ds 
\gamma_5 = \frac{i}{6} \epsilon_{\mu\rho\sigma\tau} 
\Delta^{\mu}\gamma^{\rho}
\gamma^{\sigma}\gamma^{\tau}\,.
\end{eqnarray}
Notice that this replacement is only equivalent to the HVBM prescription
if a single $\gamma_5$-matrix is present in the numerator of the OME's, 
which is the case here. Thus only one Levi-Civita
tensor appears in all the numerators of the Feynman integrals so that the 
$\gamma_5$ matrix and the Levi-Civita tensor are on an equal footing.
Since one has to be careful with the treatment
of the Levi-Civita tensor, which is also a four dimensional object, we 
have to follow the procedure in Appendix B of \cite{mene}. First one computes 
the numerator in the Feynman integral corresponding to a specific graph. 
This numerator contains the integration momenta, which are in
$N$-spacetime dimensions. Then one applies tensorial 
reduction to express the whole integral into tensors containing $\Delta_{\mu}$
in Eq. (\ref{eqn:2.4}) and $p_{\mu}$ in Eq. (\ref{eqn:2.5}),
which are $N$-dimensional vectors. Finally one has to project
the whole matrix element on the tensor structure characteristic of the 
specific OME under consideration, which follows from Lorentz covariance in 
four dimensions. Therefore the Lorentz indices of $p_\mu$ and $\Delta_\mu$ with
$N > 4$ are simply dropped.

The advantage of the HVBM prescription is that the cyclicity of the
traces is preserved. On the other hand it destroys the 
anti-commutativity of the $\gamma_5$-matrix. 
This will lead to a renormalization
of the non-singlet axial-vector current $O_{q,k}^{5,\mu}$ in 
Eq. (\ref{eqn:2.1}) in spite of the fact that
it is conserved \cite{lave}. Furthermore as was shown in 
\cite{larin} the Adler-Bardeen theorem \cite{adba} is no longer true.
The anti-commutativity and the Ward-identities can be restored by introducing
an additional renormalization constant $Z_{qq}^{5,r}$ ($r={\rm NS,S}$) which,
when expressed in unrenormalized quantities, 
has the following form
\begin{eqnarray}
\label{eqn:2.8}
\hat Z_{qq}^{5,r} &=& 1 + \hat a_s S_{\varepsilon} \Big [ z_{qq}^{(1)} 
+ \varepsilon z_{qq}^{\varepsilon,(1)}\Big ]
\nonumber\\[2ex]
&& + {\hat a}_s^2 S_{\varepsilon}^2 \Big [ -\frac{1}{\varepsilon} \beta_0
 z_{qq}^{(1)}  +  z_{qq}^{r,(2)} - 2 \beta_0  
z_{qq}^{\varepsilon,(1)} 
\nonumber\\[2ex]
&&-\hat{\xi}  \frac{d\,z_{qq}^{\varepsilon,(1)}}{d \hat{\xi}} 
z_{\xi} \Big ]_{\hat \xi = 1}\,.
\end{eqnarray}
The hat indicates that all quantities are unrenormalized with respect to 
coupling constant $\alpha_s$, gauge constant $\xi$
and, in the case of the OME's, also the operator renormalization.
Here $S_{\varepsilon}$ denotes the spherical factor characteristic of 
$N$-dimensional regularization
\begin{eqnarray}
\label{eqn:2.9}
S_{\varepsilon} = \exp\Big [ \frac{\varepsilon}{2} \Big ( \gamma_E - \ln 4\pi
\Big ) \Big ]\,,
\end{eqnarray}
where $\gamma_E$ denotes the Euler constant.
Further we introduce a shorthand notation for the strong coupling
constant so that
\begin{eqnarray}
\label{eqn:2.10}
a_s = \frac{\alpha_s}{4\pi} \,,\qquad \alpha_s = \frac{g^2}{4\pi}\,.
\end{eqnarray}
If we apply coupling constant renormalization
\begin{eqnarray}
\label{eqn:2.11}
\hat a_s = a_s \Big [ 1 + a_s S_{\varepsilon}
\Big (2\beta_0\frac{1}{\varepsilon} \Big)\Big ]\,,
\end{eqnarray}
and gauge constant renormalization
\begin{eqnarray}
\label{eqn:2.12}
\hat \xi = \xi \Big [ 1 + a_s S_{\varepsilon}
\Big (z_{\xi}\frac{1}{\varepsilon} \Big)\Big ]\,,
\end{eqnarray}
where both are presented in the ${\overline {\rm MS}}$ scheme, we obtain
\begin{eqnarray}
\label{eqn:2.13}
Z_{qq}^{5,r}&=& 1 +  a_s S_{\varepsilon} \Big [ z_{qq}^{(1)}
+ \varepsilon z_{qq}^{\varepsilon,(1)}\Big ]
\nonumber\\[2ex]
&& +  a_s^2 S_{\varepsilon}^2 \Big [ \frac{1}{\varepsilon} \beta_0
 z_{qq}^{(1)}  +  z_{qq}^{r,(2)} \Big ]\,.
\end{eqnarray}
In QCD ($SU(N)$) the renormalization group coefficients are given by
\begin{eqnarray}
\label{eqn:2.14}
\beta_0= \frac{11}{3} C_A - \frac{8}{3} n_f T_f \,,
\qquad
z_{\xi}= C_A \Big ( -\frac{10}{3} - (1- \xi) \Big ) + \frac{8}{3}n_f T_f\,,
\end{eqnarray}
with $C_F=(N^2-1)/2N$, $C_A=N$,
$T_f=1/2$ and $n_f$ stands for the number of light flavours. Notice that
the coefficients 
$z_{qq}^{(1)}$ and $z_{qq}^{r,(2)}$ 
in Eqs. (\ref{eqn:2.8}) and (\ref{eqn:2.13})
are gauge ($\xi$) independent in contrast to the 
terms proportional to $\varepsilon$
like $z_{qq}^{\varepsilon,(1)}$ which depend on the gauge.
As we will show later on $z_{qq}^{(1)}$ is universal but
$z_{qq}^{\varepsilon,(1)}$ and $z_{qq}^{r,(2)}$ depend on the given quantity.
In the non-singlet case $Z_{qq}^{5,{\rm NS}}$ 
can be inferred from the ratio
\begin{eqnarray}
\label{eqn:2.15}
Z_{qq}^{5,{\rm NS}} = \frac{\hat A_{qq}^{\rm NS,PHYS}}
{\hat A_{qq}^{5,{\rm NS,PHYS}}}\,,
\end{eqnarray}
where  $\hat A_{qq}^{\rm NS,PHYS}$ and $\hat A_{qq}^{5,{\rm NS,PHYS}}$ 
denote the physical parts (for definitions see below) of the OME's
without and with the $\gamma_5$-matrix respectively. The former OME has been
fully calculated in \cite{msn} whereas the latter OME will be presented below
and in Appendix A. Notice that the numerator and the denominator would
be identical if we applied the so-called naive
(or anticommuting) $\gamma_5$-prescription. 
The latter method allows us to anticommute this matrix with the other
$\gamma_\mu$-matrices so that it 
reaches another vertex in the diagram and leads to  
$Z_{qq}^{5,{\rm NS}}=1$. Therefore the 
renormalization constant $Z_{qq}^{5,{\rm NS}}$ restores the anti-commutativity
of the $\gamma_5$-matrix. The singlet coefficient which can be split into
\begin{eqnarray}
\label{eqn:2.16}
z_{qq}^{{\rm S},(2)} = z_{qq}^{{\rm NS},(2)} + z_{qq}^{{\rm PS},(2)}
\,,
\end{eqnarray} 
can also be calculated by anticommuting the $\gamma_5$-matrix. 
The computation of the purely
singlet part $z_{qq}^{{\rm PS},(2)}$ will be presented at the end of this 
section.

To check our results for the sums of the Feynman graphs it is useful
to have explicit expressions for the pole terms. Therefore we 
will now present the OME's expressed in renormalization group 
coefficients defined in \cite{mene}.
The explicit formulae for the OME's can be found in the Appendix A and
contain terms which are finite in the limit $\varepsilon\rightarrow 0$.
Further
it is implicitly understood that all quantities in this section, in 
particular the anomalous dimensions $\gamma_{ij}$ ($i,j=q,g$), are 
Mellin transforms even though we suppress the moment index $n$ to simplify
the notation. 
(Another way to interpret the formulae is that
the OME's are given in parton momentum fraction ($z$) space when the anomalous
dimensions are replaced by minus the corresponding 
Altarelli-Parisi splitting functions and the multiplications are replaced
by convolutions.)
We have written the OME's in such a way that all 
renormalization group coefficients appearing in the expressions below are
given in the ${\overline {\rm MS}}$-scheme.

In the rest of the section the superscript 5 will be suppressed because 
we will only discuss the spin OME's.
Up to order $\alpha_s^2$ the non-singlet and the singlet OME's 
can be decomposed into
\begin{eqnarray}
\label{eqn:2.17}
\hat A_{qq}^r= \Big [ \gamma_5 
{\Delta \hspace{-0.52em}/\hspace{0.1em}}
%\Ds 
\hat A_{qq}^{r,{\rm PHYS}} + \gamma_5 
{p \hspace{-0.52em}/\hspace{0.1em}}
%\ps 
\frac{\Delta\cdot p}{p^2} \hat A_{qq}^{r,{\rm EOM}} \Big ]
(\Delta\cdot p)^{n-1}\,,
\end{eqnarray}
where $A_{qq}^{r,{\rm PHYS}}$ 
and $A_{qq}^{r,{\rm EOM}}$ stand for the physical and 
unphysical parts respectively with $r={\rm NS,S}$. The presence of the
latter is due to the fact that the equations of motion are not satisfied. 
The non-singlet physical OME can now be expressed in renormalization 
group coefficients as follows
\begin{eqnarray}
\label{eqn:2.18}
\hat A_{qq}^{\rm NS,PHYS} &=& 1 + \hat a_s S_{\varepsilon} 
\big(\frac{-p^2}{\mu^2}\big)^{\varepsilon/2}
\Big [ \frac{1}{\varepsilon} \gamma_{qq}^{{\rm NS},(0)} + a_{qq}^{{\rm NS},(1)}
- z_{qq}^{(1)} 
\nonumber\\[2ex]
&& +\varepsilon ( a_{qq}^{{\rm NS},\varepsilon,(1)}
-z_{qq}^{\varepsilon,(1)}) \Big ]
\nonumber\\[2ex]
&& +{\hat a}_s^2 S_{\varepsilon}^2 
\big(\frac{-p^2}{\mu^2}\big)^{\varepsilon} \Big[
 \frac{1}{\varepsilon^2} \Big \{ \frac{1}{2} (\gamma_{qq}^{{\rm NS},(0)})^2
 - \beta_0 \gamma_{qq}^{{\rm NS},(0)} \Big \}
\nonumber\\[2ex]
&& + \frac{1}{\varepsilon}\Big \{ \frac{1}{2} \gamma_{qq}^{{\rm NS},(1)}
 - 2 \beta_0 z_{qq}^{(1)}
- 2 \beta_0 \Big (a_{qq}^{{\rm NS},(1)} - z_{qq}^{(1)} \Big )
\nonumber\\[2ex]
&& + \gamma_{qq}^{{\rm NS},(0)} \Big ( a_{qq}^{{\rm NS},(1)} 
- z_{qq}^{(1)} \Big )
- \hat{\xi}  \frac{d\,\Big (a_{qq}^{{\rm NS},(1)} - z_{qq}^{(1)} \Big )}
{d \hat{\xi}}  z_{\xi} \Big \}
\nonumber\\[2ex]
&& + a_{qq}^{{\rm NS},(2)}- z_{qq}^{{\rm NS},(2)} - z_{qq}^{(1)} \Big (
a_{qq}^{{\rm NS},(1)} -z_{qq}^{(1)} \Big )
\nonumber\\[2ex]
&& - 2 \beta_0 \Big ( a_{qq}^{{\rm NS},\varepsilon,(1)}
- z_{qq}^{\varepsilon,(1)} \Big )
+ \gamma_{qq}^{{\rm NS},(0)} \Big ( a_{qq}^{{\rm NS},\varepsilon,(1)}
- z_{qq}^{\varepsilon,(1)} \Big )
\nonumber\\[2ex]
&& - \hat{\xi}  \frac{d\,\Big (a_{qq}^{{\rm NS},\varepsilon,(1)}
- z_{qq}^{\varepsilon,(1)} \Big )}{d \hat{\xi}} z_{\xi}
\Big ]_{\hat{\xi}=1}\,.
\end{eqnarray}
The $\gamma_{ij}^{(k)}$ denote the coefficients of the 
order $a_s^{k+1}$ terms appearing in the series expansions for the 
anomalous dimensions. Using the same notation we can also express 
the unphysical part of the non-singlet OME 
in Eq. (\ref{eqn:2.17}) in the aforementioned renormalization 
group coefficients
\begin{eqnarray}
\label{eqn:2.19}
\hat A_{qq}^{\rm NS,EOM}&=&  \hat a_s S_{\varepsilon} 
\big(\frac{-p^2}{\mu^2}\big)
^{\varepsilon/2} \Big [  b_{qq}^{{\rm NS},(1)}
+ \varepsilon b_{qq}^{{\rm NS},\varepsilon,(1)}\Big ]
\nonumber\\[2ex]
&& +{\hat a}_s^2 S_{\varepsilon}^2 
\big(\frac{-p^2}{\mu^2}\big)^{\varepsilon} \Big[
\frac{1}{\varepsilon} \Big \{ \gamma_{qq}^{{\rm NS},(0)} b_{qq}^{{\rm NS},(1)}
- 2 \beta_0 b_{qq}^{{\rm NS},(1)} 
\nonumber\\[2ex]
&& - \hat{\xi}  \frac{d~b_{qq}^{{\rm NS},(1)}}
{d \hat{\xi} } z_{\xi} \Big \}
+ b_{qq}^{{\rm NS},(2)} - 2 \beta_0 b_{qq}^{{\rm NS},\varepsilon,(1)}
+ \gamma_{qq}^{{\rm NS},(0)} b_{qq}^{{\rm NS},\varepsilon,(1)}
\nonumber\\[2ex]
&& - z_{qq}^{(1)}  b_{qq}^{{\rm NS},(1)} 
- \hat{\xi}  \frac{d~b_{qq}^{{\rm NS},\varepsilon,(1)}} {d \hat{\xi} }
z_{\xi} \Big ]_{\hat{\xi}=1} \,.
\end{eqnarray}
The singlet OME can be decomposed as follows
\begin{eqnarray}
\label{eqn:2.20}
\hat A_{qq}^{\rm S}= \hat A_{qq}^{\rm NS} + \hat A_{qq}^{\rm PS} \,,
\end{eqnarray}
where the purely singlet physical part equals
\begin{eqnarray}
\label{eqn:2.21}
\hat A_{qq}^{\rm PS, PHYS} &=&
 {\hat a}_s^2 S_{\varepsilon}^2 (\frac{-p^2}{\mu^2})^{\varepsilon} \Big[
 \frac{1}{\varepsilon^2} \Big \{ \frac{1}{2} \gamma_{qg}^{(0)}
\gamma_{gq}^{(0)} \Big \}
\nonumber\\[2ex]
&& + \frac{1}{\varepsilon}\Big \{ \frac{1}{2} \gamma_{qq}^{{\rm PS},(1)}
+ \gamma_{qg}^{(0)} a_{gq}^{(1)} \Big \}
+ a_{qq}^{{\rm PS},(2)} - z_{qq}^{{\rm PS},(2)} 
\nonumber\\[2ex]
&& + \gamma_{qg}^{(0)} a_{gq}^{\varepsilon,(1)}\Big ]\,,
\end{eqnarray}
and the purely singlet unphysical part equals
\begin{eqnarray}
\label{eqn:2.22}
\hat A_{qq}^{\rm PS,EOM}&=&
{\hat a}_s^2 S_{\varepsilon}^2 (\frac{-p^2}{\mu^2})^{\varepsilon} \Big[
\frac{1}{\varepsilon} \Big \{ \gamma_{qg}^{(0)} b_{gq}^{(1)} \Big \}
+ b_{qq}^{{\rm PS},(2)} + \gamma_{qg}^{(0)} b_{gq}^{\varepsilon,(1)} 
\Big ] \,.
\end{eqnarray}
The next OME is 
\begin{eqnarray}
\label{eqn:2.23}
\hat A_{qg,\mu\nu}=
\epsilon_{\mu\nu\alpha\beta}\Delta^{\alpha}p^{\beta}
\frac{1}{\Delta \cdot p} 
\hat A_{qg}^{\rm PHYS} \,, 
\end{eqnarray}
with
\begin{eqnarray}
\label{eqn:2.24}
\hat A_{qg}^{\rm PHYS} &=&  \hat a_s S_{\varepsilon} 
\big(\frac{-p^2}{\mu^2}\big)
^{\varepsilon/2} \Big [ \frac{1}{\varepsilon} \gamma_{qg}^{(0)} + a_{qg}^{(1)}
+ \varepsilon a_{qg}^{\varepsilon,(1)}\Big ]
\nonumber\\[2ex]
&&+{\hat a}_s^2 S_{\varepsilon}^2 
\big(\frac{-p^2}{\mu^2}\big)^{\varepsilon} \Big[
 \frac{1}{\varepsilon^2} \Big \{ \frac{1}{2} 
\Big (\gamma^{(0)}_{qq} \gamma_{qg}^{(0)} 
+\gamma_{qg}^{(0)}\gamma_{gg}^{(0)} \Big )
- \beta_0 \gamma_{qg}^{(0)} \Big \}
\nonumber\\[2ex]
&& +\frac{1}{\varepsilon} \Big \{ \frac{1}{2} \Big ( \gamma_{qg}^{(1)}
- z_{qq}^{(1)} \gamma_{qg}^{(0)}  \Big ) - 2 \beta_0 a_{qg}^{(1)} +
\gamma_{qg}^{(0)} a_{gg}^{(1)} + \gamma_{qq}^{(0)} a_{qg}^{(1)}
\nonumber\\[2ex]
&& - \hat{\xi}  \frac{d~a_{qg}^{(1)}}
{d \hat{\xi}} z_{\xi} \Big \}
+ a_{qg}^{(2)} - 2 \beta_0 a_{qg}^{\varepsilon,(1)}
+ \gamma_{qq}^{(0)} a_{qg}^{\varepsilon,(1)}
+ \gamma_{qg}^{(0)} a_{gg}^{\varepsilon,(1)}
\nonumber\\[2ex]
&& - z_{qq}^{(1)} a_{qg}^{(1)} - z_{qq}^{\varepsilon,(1)} \gamma_{qg}^{(0)}
- \hat{\xi}  \frac{d~a_{qg}^{\varepsilon,(1)}}{d \hat{\xi}} z_{\xi}
\Big ]_{\hat{\xi}=1} \,.
\end{eqnarray}
Next we need
\begin{eqnarray}
\label{eqn:2.25}
\hat A_{gq}= \Big [\gamma_5 
{\Delta \hspace{-0.52em}/\hspace{0.1em}}
%\Ds 
\hat A_{gq}^{\rm PHYS} + 
\gamma_5 {p \hspace{-0.52em}/\hspace{0.1em}}
%\ps 
\frac{\Delta \cdot p}{p^2} 
\hat A_{gq}^{\rm EOM} \Big ] (\Delta \cdot p)^{n-1}\,,
\end{eqnarray}
where
\begin{eqnarray}
\label{eqn:2.26}
\hat A_{gq}^{\rm PHYS} &=& \hat a_s S_{\varepsilon} (\frac{-p^2}{\mu^2})
^{\varepsilon/2}
\Big [ \frac{1}{\varepsilon} \gamma_{gq}^{(0)} + a_{gq}^{(1)}
+ \varepsilon a_{gq}^{\varepsilon,(1)}  \Big ]
\nonumber\\[2ex]
&& +{\hat a}_s^2 S_{\varepsilon}^2 (\frac{-p^2}{\mu^2})^{\varepsilon} \Big[
 \frac{1}{\varepsilon^2} \Big \{ \frac{1}{2} ( \gamma_{gg}^{(0)}
 \gamma_{gq}^{(0)} + \gamma^{(0)}_{gq} \gamma_{qq}^{(0)} )
 - \beta_0 \gamma_{gq}^{(0)} \Big \}
\nonumber\\[2ex]
&& + \frac{1}{\varepsilon}\Big \{ \frac{1}{2} \Big ( \gamma_{gq}^{(1)}
+ \gamma_{gq}^{(0)} z_{qq}^{(1)} \Big ) - 2 \beta_0 a_{gq}^{(1)} +
\gamma_{gq}^{(0)} \Big ( a_{qq}^{(1)} - z_{qq}^{(1)} \Big )
\nonumber\\[2ex]
&& + \gamma_{gg}^{(0)} a_{gq}^{(1)}
 - \hat{\xi}  \frac{d~a_{gq}^{(1)}}
{d \hat{\xi} } z_{\xi} \Big \} + a_{gq}^{(2)}
- 2 \beta_0 a_{gq}^{\varepsilon,(1)}+\gamma_{gq}^{(0)} z_{qq}^{\varepsilon,(1)}
\nonumber\\[2ex]
&& + \gamma_{gq}^{(0)}\Big (  a_{qq}^{\varepsilon,(1)} - 
z_{qq}^{\varepsilon, (1)}\Big ) + \gamma_{gg}^{(0)} a_{gq}^{\varepsilon, (1)}
 - \hat{\xi}  \frac{d~a_{gq}^{\varepsilon,(1)}}{d \hat{\xi} } z_{\xi}
\Big ]_{\hat{\xi}=1} \,,
\end{eqnarray}
and
\begin{eqnarray}
\label{eqn:2.27}
\hat A_{gq}^{\rm EOM}&=&  \hat a_s S_{\varepsilon} (\frac{-p^2}{\mu^2})
^{\varepsilon/2} \Big [  b_{gq}^{(1)}
+ \varepsilon b_{gq}^{\varepsilon,(1)}\Big ]
\nonumber\\[2ex]
&& +{\hat a}_s^2 S_{\varepsilon}^2 (\frac{-p^2}{\mu^2})^{\varepsilon} \Big[
\frac{1}{\varepsilon} \Big \{  \gamma_{gq}^{(0)}  b_{qq}^{(1)}
+ \gamma_{gg}^{(0)}  b_{gq}^{(1)}- 2 \beta_0 b_{gq}^{(1)} 
\nonumber\\[2ex]
&& - \hat{\xi}  \frac{d~b_{gq}^{(1)}}
{d \hat{\xi} } z_{\xi} \Big \} + b_{gq}^{(2)} -2 \beta_0 
b_{gq}^{\varepsilon,(1)}
+ \gamma_{gq}^{(0)}  b_{qq}^{{\rm NS},\varepsilon,(1)}+ \gamma_{gg}^{(0)}  
b_{gq}^{\varepsilon,(1)}
\nonumber\\[2ex]
&& - \hat{\xi}\frac{d~b_{gq}^{{\rm NS},\varepsilon,(1)}}{d \hat{\xi} } z_{\xi}
\Big ]_{\hat{\xi}=1}\,.
\end{eqnarray}
Finally we need
\begin{eqnarray}
\label{eqn:2.28}
\hat A_{gg,\mu\nu}=
\epsilon_{\mu\nu\alpha\beta}\Delta^{\alpha}p^{\beta} 
\frac{1}{\Delta \cdot p} 
\hat A_{gg}^{\rm PHYS} \,,
\end{eqnarray}
with
\begin{eqnarray}
\label{eqn:2.29}
\hat A_{gg}^{\rm PHYS}&=& 1 + \hat a_s S_{\varepsilon} (\frac{-p^2}{\mu^2})
^{\varepsilon/2} \Big [ \frac{1}{\varepsilon} \gamma_{gg}^{(0)} + a_{gg}^{(1)}
+ \varepsilon a_{gg}^{\varepsilon,(1)}\Big ]
\nonumber\\[2ex]
&&+{\hat a}_s^2 S_{\varepsilon}^2 (\frac{-p^2}{\mu^2})^{\varepsilon} \Big[
 \frac{1}{\varepsilon^2} \Big \{ \frac{1}{2} \Big ((\gamma_{gg}^{(0)})^2
+ \gamma_{gq}^{(0)} \gamma_{qg}^{(0)} \Big )
- \beta_0 \gamma_{gg}^{(0)} \Big \}
\nonumber\\[2ex]
&& +\frac{1}{\varepsilon} \Big \{ \frac{1}{2} \gamma_{gg}^{(1)}
- 2 \beta_0 a_{gg}^{(1)} +
\gamma_{gg}^{(0)} a_{gg}^{(1)} + \gamma_{gq}^{(0)} a_{qg}^{(1)}
- \hat{\xi}  \frac{d~a_{gg}^{(1)}} {d \hat{\xi} } z_{\xi} \Big \}
\nonumber\\[2ex]
&& + a_{gg}^{(2)} - 2 \beta_0 a_{gg}^{\varepsilon,(1)} + \gamma_{gg}^{(0)}
a_{gg}^{\varepsilon,(1)}+ \gamma_{gq}^{(0)} a_{qg}^{\varepsilon,(1)}
- \hat{\xi}  \frac{d~a_{gg}^{\varepsilon,(1)}} {d \hat{\xi} } z_{\xi}
\Big ]_{\hat{\xi}=1}\,.
\end{eqnarray}

Explicit formulae for the above unrenormalized OME's corrected up to second
order, which contain all the finite terms independent of $\varepsilon$, 
are given in Appendix A. The pole terms agree with the results
in Eqs. (2.18)-(2.29). Since the 
external quark and gluon legs are off-shell one can directly insert 
the formulae in Appendix A into Feynman integrals for one-loop graphs. 
In this way one gets results for subsets of the three-loop graphs. 
It is however 
clear that the most difficult Feynman integrals containing non-planar
diagrams, where all quark or gluon lines cross, remain to be done.

The renormalized OME's are given by
\begin{eqnarray}
\label{eqn:2.30}
A_{qq}^{\rm NS} = Z_{qq}^{5,{\rm NS}} (Z^{-1})_{qq}^{\rm NS} 
\hat A_{qq}^{\rm NS} \,,
\end{eqnarray}
\begin{eqnarray}
\label{eqn:2.31}
A_{qq} = Z_{qq}^{5,{\rm S}} (Z^{-1})_{qq}^{\rm S} \hat A_{qq} + (Z^{-1})_{qg}
\hat A_{gq}\,,
\end{eqnarray}
\begin{eqnarray}
\label{eqn:2.32}
A_{qg} = Z_{qq}^{5,{\rm S}} (Z^{-1})_{qq}^{\rm S} \hat A_{qg} + (Z^{-1})_{qg}
\hat A_{gg}\,,
\end{eqnarray}
\begin{eqnarray}
\label{eqn:2.33}
A_{gq} = Z_{qq}^{5,{\rm S}} (Z^{-1})_{gq} \hat A_{qq} + (Z^{-1})_{gg}
\hat A_{gq}\,,
\end{eqnarray}
and
\begin{eqnarray}
\label{eqn:2.34}
A_{gg} = Z_{qq}^{5,{\rm S}} (Z^{-1})_{gq} \hat A_{qg} + (Z^{-1})_{gg}
\hat A_{gg}\,.
\end{eqnarray}
The operator renormalization constants are given by 
\begin{eqnarray}
\label{eqn:2.35}
(Z^{-1})_{qq}^{\rm S}= (Z^{-1})_{qq}^{\rm NS}+ (Z^{-1})_{qq}^{\rm PS}\,,
\end{eqnarray}
where
\begin{eqnarray}
\label{eqn:2.36}
(Z^{-1})_{qq}^{\rm NS}&=& 1 + a_s S_{\varepsilon} \Big [-\frac{1}{\varepsilon}
\gamma_{qq}^{(0)} \Big ] + a_s^2 S_{\varepsilon}^2 \Big [
\frac{1}{\varepsilon^2}\Big \{
\frac{1}{2}  (\gamma_{qq}^{(0)})^2 
- \beta_0 \gamma_{qq}^{(0)} \Big \}
\nonumber\\[2ex]
&& -  \frac{1}{2\varepsilon}  \gamma_{qq}^{{\rm NS},(1)}
 \Big ] \,,
\end{eqnarray}
and
\begin{eqnarray}
\label{eqn:2.37}
(Z^{-1})_{qq}^{\rm PS}&=&
 a_s^2 S_{\varepsilon}^2 \Big [ \frac{1}{\varepsilon^2} \Big \{\frac{1}{2}
\gamma_{qg}^{(0)} \gamma_{gq}^{(0)} \Big \} -  \frac{1}{2\varepsilon}
\gamma_{qq}^{{\rm PS},(1)}  \Big ] \,.
\end{eqnarray}
The other singlet operator renormalization constants are given by
\begin{eqnarray}
\label{eqn:2.38}
(Z^{-1})_{qg}&=&  a_s S_{\varepsilon} \Big [-\frac{1}{\varepsilon}
\gamma_{qg}^{(0)} \Big ] + a_s^2 S_{\varepsilon}^2 \Big [
\frac{1}{\varepsilon^2}\Big \{
\frac{1}{2}  \Big ( \gamma^{(0)}_{qq} \gamma_{qg}^{(0)} 
+ \gamma_{qg}^{(0)} \gamma_{gg}^{(0)}
 \Big ) - \beta_0 \gamma_{qg}^{(0)} \Big \}
\nonumber\\[2ex]
&& - \frac{1}{2 \varepsilon} \Big \{  \gamma_{qg}^{(1)}
+ z_{qq}^{(1)} \gamma_{qg}^{(0)}  \Big \}  \Big ]\,,
\end{eqnarray}
\begin{eqnarray}
\label{eqn:2.39}
(Z^{-1})_{gq}&=&  a_s S_{\varepsilon} \Big [-\frac{1}{\varepsilon}
\gamma_{gq}^{(0)} \Big ] + a_s^2 S_{\varepsilon}^2 \Big [
\frac{1}{\varepsilon^2}\Big \{
\frac{1}{2}  \Big ( \gamma^{(0)}_{gq} \gamma_{qq}^{(0)} +
\gamma_{gg}^{(0)} \gamma_{gq}^{(0)}
\Big ) - \beta_0 \gamma_{gq}^{(0)} \Big \}
\nonumber\\[2ex]
&& - \frac{1}{2 \varepsilon} \Big \{ \gamma_{gq}^{(1)}
- z_{qq}^{(1)} \gamma_{gq}^{(0)} \Big \}   \Big ]\,,
\end{eqnarray}
and
\begin{eqnarray}
\label{eqn:2.40}
(Z^{-1})_{gg}&=& 1 + a_s S_{\varepsilon} \Big [-\frac{1}{\varepsilon}
\gamma_{gg}^{(0)} \Big ] + a_s^2 S_{\varepsilon}^2 \Big [
\frac{1}{\varepsilon^2}\Big \{
\frac{1}{2} \Big ( (\gamma_{gg}^{(0)})^2 + \gamma_{gq}^{(0)}
\gamma_{qg}^{(0)} \Big ) - \beta_0 \gamma_{gg}^{(0)} \Big \}
\nonumber\\[2ex]
&& - \frac{1}{2 \varepsilon} \gamma_{gg}^{(1)} \Big ]\,.
\end{eqnarray}
The above renormalization constants are chosen in such a way that the
anomalous dimensions $\gamma_{ij}^{(1)}$ are presented in the
${\overline {\rm MS}}$-scheme. Naively one would expect that the single
pole terms in Eqs. (\ref{eqn:2.38}) and (\ref{eqn:2.39}) are then given by
$-\gamma_{qg}^{(1)}/2$ and $-\gamma_{gq}^{(1)}/2$ respectively. However
due to the HVBM prescription the renormalization constants $Z_{ij}$ have to
be modified in order to bring the renormalized OME's in the standard form.
By this we mean that  
the latter satisfy the renormalization group equations with the 
anomalous dimensions $\gamma_{ij}^{(k)}$. This we will leave as
a check for the reader. Notice that only the final OME's are independent
of the prescription for the $\gamma_5$-matrix and the Levi-Civita tensor. 

To conclude we discuss properties of the renormalization 
factors $Z_{qq}^{5,r}$, $r=$ NS, S in (\ref{eqn:2.8}) computed in this paper 
and compare them with other results in the literature. 
Starting with the non-singlet term, which is computed via
(\ref{eqn:2.15}), the unrenormalized expression can be  
found in Eq. (\ref{eqn:A.11}). Due to the factor $(-1)^n$ 
(for its origin see below Eq. (\ref{eqn:A.3})) it can be split into two parts
as follows
\begin{eqnarray}
\label{eqn:2.41}
Z_{qq}^{5,{\rm NS}} = \int_0^1 \,dz\, z^{n-1} 
\Big [ Z_{qq}^{5,{\rm NS},(+)} (z)
+ (-1)^n Z_{qq}^{5,{\rm NS},(-)} (z) \Big ]\,.
\end{eqnarray}
After coupling constant and gauge constant renormalization (see 
Eqs. (\ref{eqn:2.11}), (\ref{eqn:2.12})) the two parts become
\begin{eqnarray}
\label{eqn:2.42}
Z_{qq}^{5,{\rm NS},(+)} (z) &=&  \delta(1-z) 
+  a_s S_{\varepsilon} C_F \Big [ - 8 (1-z) 
+ \varepsilon \Big \{ - 4(1-z)\ln(1-z)
\nonumber\\[2ex]
&& -4(1 - z)\ln z + 2z -(1-\xi)  \Big \} \Big ]
\nonumber\\[2ex]
&& +  a_s^2 S_{\varepsilon}^2 \Big [ C_F^2 \Big \{
 - 16 (1 - z) - (16 + 8 z)\ln z 
\nonumber\\[2ex]
&& + 16(1-z)\ln z \ln(1-z)  \Big \}
\nonumber\\[2ex]
&& + C_A C_F \Big \{ - \frac{1}{\varepsilon} \Big ( \frac{88}{3}(1-z) \Big )
-\frac{592}{9}(1 - z) + 8(1-z)\zeta(2)
\nonumber\\[2ex]
&& +(- \frac{80}{3}+\frac{8}{3}z)\ln z -4(1-z)\ln^2 z \Big \}
\nonumber\\[2ex]
&& + n_f C_F T_f \Big \{
\frac{1}{\varepsilon}\Big ( \frac{32}{3}(1-z) \Big ) + \frac{80}{9} (1 - z)
\nonumber\\[2ex]
&& + \frac{16}{3}(1-z)\ln z \Big \} \Big ]\,,
\end{eqnarray}
and
\begin{eqnarray}
\label{eqn:2.43}
Z_{qq}^{5,{\rm NS},(-)} (z) &=& a_s^2 S_{\varepsilon}^2 \Big [ 
(C_F^2-\frac{1}{2}C_A C_F) \Big \{  8(1+z) \Big ( 4 {\rm Li}_2(-z)
+ 4\ln z \ln (1+z)
\nonumber\\[2ex]
&& + 2\zeta(2) - \ln^2 z - 3\ln z \Big ) - 56(1-z) \Big \} \Big ] \,,
\end{eqnarray}
respectively. Choosing  $n=1$ in Eq. (\ref{eqn:2.41}) we obtain
\begin{eqnarray}
\label{eqn:2.44}
Z_{qq}^{5,{\rm NS}} & = & 1 + a_s S_{\varepsilon} C_F \Big [ - 4 + 
 \varepsilon \Big (5 - (1-\xi) \Big )\Big ]
\nonumber\\[2ex]
&& + a_s^2 S_{\varepsilon}^2 \Big [ C_F^2 \Big \{ 22  \Big \}
+ C_A C_F \Big \{
- \frac{44}{3} \frac{1}{\varepsilon} - \frac{107}{9} \Big \}
\nonumber\\[2ex]
&& + n_f C_F T_f
\Big \{ \frac{16}{3} \frac{1}{\varepsilon} + \frac{4}{9} \Big \} \Big ]\,.
\end{eqnarray}
This is the renormalization constant which has to be
multiplied with the non-singlet axial-vector current 
$O_{q,k}^{5,\mu}$ in Eq. (\ref{eqn:2.1})
so that the latter will not be renormalized (i.e., remain a conserved
current) if one employs the HVBM prescription. 
This renormalization constant has been calculated up to order $\alpha_s^3$
in \cite{larin}, \cite{lave}. Except for the term proportional to
$\varepsilon$, which was not presented previously, 
our result in Eq. (\ref{eqn:2.44}) 
agrees with that in the aforementioned literature. Recently the factor 
$Z_{qq}^{5,{\rm NS}}$
has been also computed in \cite{rijk} for the coefficient functions
corresponding to the longitudinal spin fragmentation function 
in ${\rm e}^ + {\rm e}^-$-scattering. 
If we indicate the latter factor by the superscript T (timelike) and the one
in Eq. (\ref{eqn:2.41}) by S (spacelike) we have the following relation
\begin{eqnarray}
\label{eqn:2.45}
Z_{qq}^{5,{\rm NS},(+),{\rm T}} (z) = -z Z_{qq}^{5,{\rm NS},(+),{\rm S}}
(\frac{1}{z}) + a_s^2 \Big [\beta_0 z_{qq}^{{\rm NS},(1)}(z) \ln z\Big ]\,.
\end{eqnarray}
This result demonstrates the breakdown of the Gribov-Lipatov relation
\cite{grli} which is therefore only valid up to order $\alpha_s$. 
However for the minus component 
\begin{eqnarray}
\label{eqn:2.46}
Z_{qq}^{5,{\rm NS},(-),{\rm T}} (z)=-z Z_{qq}^{5,{\rm NS},(-),{\rm S}}
(\frac{1}{z}) \,,
\end{eqnarray}
so the Gribov-Lipatov relation still holds at least up to second order.
The above relations have also been reported for the
spin averaged non-singlet spacelike (S) and timelike (T) splitting functions 
$P_{qq}^{\rm NS}$ in Eqs. (6.37) and (6.38) of \cite{fupe} where 
$z_{qq}^{(1)}$ is replaced by $P_{qq}^{(0)}$. Notice that there
is no difference between spin and spin averaged splitting functions as far
as the non-singlet part is concerned (see \cite{stvo}). However there is 
a difference between $Z_{qq}^{5,{\rm NS}}$ and $P_{qq}^{\rm NS}$ if one looks
at relation (6.46) in \cite{fupe}. In our case it reads
\begin{eqnarray}
\label{eqn:2.47}
Z_{qq}^{5,{\rm NS},(+),{\rm T}}(z)= Z_{qq}^{5,{\rm NS},(+),{\rm S}}(z) + a_s^2
\frac{1}{2}\ln z \int_z^1\,dy\,\Big ( z_{qq}^{{\rm NS},(1)}(\frac{z}{y}) 
 P_{qq}^{(0)}(y) \Big )\,,
\end{eqnarray}
whereas in Eq. (6.46) the $\ln z$ is shifted to the integrand where it becomes
$\ln y$. Hence the first moments of $Z_{qq}^{5,{\rm NS,T}}$ and 
$Z_{qq}^{5,{\rm NS,S}}$ are not equal anymore. It implies that the 
coefficient 22 of the $C_F^2$ part of $Z_{qq}^{5,{\rm NS,S}}$  
in Eq. (\ref{eqn:2.44}) is changed into $12 + 16 \zeta(2)$ 
for $Z_{qq}^{5,{\rm NS,T}}$ (see Eq. (3.46) in \cite{rijk}).

The calculation of $Z_{qq}^{{\rm PS},(2)}$, which can
only come from $\hat A_{qq}^{\rm PS,PHYS}$ in Eq. (\ref{eqn:2.21}), proceeds 
as follows.
Up to second order only one diagram contributes to this OME, which can
be found in Fig. 3.1 of \cite{mene}.
Following the same arguments as given below 
Eq. (\ref{eqn:2.15}) one has to anticommute the $\gamma_5$-matrix to another 
vertex in the diagram so that $Z_{qq}^{5,{\rm PS}}=1$. However in this 
procedure the Adler-Bell-Jackiw (ABJ) anomaly, characteristic of this graph, 
will also then be removed. In this case the 
expression for Eq. (\ref{eqn:2.21}) becomes
\begin{eqnarray}
\label{eqn:2.48}
\hat A_{qq}^{\rm PS, PHYS,ABJ} &=&
 {\hat a}_s^2 S_{\varepsilon}^2 (\frac{-p^2}{\mu^2})^{\varepsilon} \Big[
 \frac{1}{\varepsilon^2} \Big \{ \frac{1}{2} \gamma_{qg}^{(0)}
\gamma_{gq}^{(0)} \Big \}
\nonumber\\[2ex]
&& + \frac{1}{\varepsilon}\Big \{ \frac{1}{2} \Big (\gamma_{qq}^{{\rm PS},(1)}
+ v_{qg}^{(1)}\gamma_{gq}^{(0)}\Big ) + \gamma_{qg}^{(0)} a_{gq}^{(1)} \Big \}
+ a_{qq}^{{\rm PS},(2)}
\nonumber\\[2ex]
&& + v_{qg}^{(1)} a_{gq}^{(1)} + \gamma_{qg}^{(0)} a_{gq}^{\varepsilon,(1)}
\Big ]\,.
\end{eqnarray}
Comparing the above expression with the one in Eq. (\ref{eqn:2.21}) we observe
that in the former the effective anomalous dimension has become
$\gamma_{qq}^{{\rm PS},(1)} + v_{qg}^{(1)}\gamma_{gq}^{(0)}$ with 
$v_{qg}^{(1)}=8 T_f (1-z)$.
If we take the first moment of this expression this effective anomalous 
dimension becomes zero meaning that the ABJ-anomaly vanishes. 
Therefore there is a 
striking analogy between $v_{qg}^{(1)}$ above and $z_{qq}^{(1)}$ in Eq. 
(\ref{eqn:2.8}). Both constants remove anomalies i.e. the former in the case 
of the singlet current $O_q^{5,\mu}$ and the latter in the case of  
the non-singlet current $O_{q,k}^{5,\mu}$. 
However the ABJ anomaly has to be restored. Hence 
from Eqs. 
(\ref{eqn:2.21}), (\ref{eqn:2.26}) and (\ref{eqn:2.48}), 
we get \begin{eqnarray}
\label{eqn:2.49}
&& \hat A_{qq}^{\rm PS, PHYS,ABJ}-\hat A_{qq}^{\rm PS, PHYS} 
- {\hat a}_s S_{\varepsilon} \Big [v_{qg}^{(1)} \Big ] A_{gq}^{\rm PHYS} + 
{\hat a}_s^2 S_{\varepsilon}^2 
\Big [ \frac{1}{2\varepsilon}\Big (v_{qg}^{(1)} \gamma_{gq}^{(0)}\Big )
\Big ] 
\nonumber\\[2ex]
&&= {\hat a}_s^2 S_{\varepsilon}^2 \Big [ z_{qq}^{{\rm PS},(2)} \Big ]\,,
\end{eqnarray}
where the last term before the equal sign restores the ABJ anomaly 
in $\hat A_{qq}^{\rm PS, PHYS,ABJ}$. Finally we obtain
\begin{eqnarray}
\label{eqn:2.50}
z_{qq}^{{\rm PS},(2)}= n_f C_F T_f \Big [ 16(1-z) + 8 (3-z)\ln z 
+ 4 (2+z)\ln^2 z \Big ]\,.
\end{eqnarray}
If one takes the first moment of this expression the result becomes
$3 n_f C_F T_f$ which agrees with the one quoted in \cite{larin}.

Finally one could ask why it is preferable to choose the HVBM instead of
the naive $\gamma_5$-prescription since in the latter case
$Z_{qq}^{5,r} = 1$. The first reason is that HVBM was originally proposed to
obtain the ABJ-anomaly in the fermion triangle graph, which
is not always correctly reproduced using the naive prescription
(see (\ref{eqn:2.48})).  The second reason is that the Levi-Civita 
tensor appearing in the OME $A_{gq}$ induces the HVBM prescription in the 
subgraphs containing quark loops. Therefore the naive $\gamma_5$-prescription
does not prevent an additional renormalization so it is inconsistent. It is 
better to use a consistent procedure like the HVBM where all constants 
are fixed once and for all.
%%%%%%%%%%%%%%%%%%%%%%%%% ACKNOWLEDGEMENTS %%%%%%%%%%%%%%%%%%%%%%%%%%%%%%%%%%%
%
%\newpage
\acknowledgments
This research was supported in part 
by the National Science Foundation grant PHY-9722101.
%%%%% APPENDICES %%%%%%%%%%%%%%%%%%%%%%
\appendix 
\section{}
%\appendix{Appendix A}

In this Appendix we present complete expressions for the two-loop corrected 
OME's computed from the Feynman diagrams depicted
in \cite{mene}. The second order contributions are
calculated up to finite terms which survive in the limit
$\varepsilon \rightarrow 0$.
The OME's presented here are unrenormalized and external
self-energy corrections are included.
In these expressions definitions of the Riemann zeta-functions
$\zeta(n)$ and the polylogarithms ${\rm Li}_n(z)$, ${\rm S}_{n,m}(z)$ can be
found in \cite{lbmr}. Also the distributions
$(1/(1-z))_+$ and $(\ln(1-z)/(1-z))_+$ and written as
$1/(1-z)$ and $\ln(1-z)/(1-z)$ respectively to shorten the formulae.
Note that the OME's given in the text are the
moments of the functions listed here so
%(A.1)
\begin{eqnarray}
\label{eqn:A.1}
A_{ij}^n = \int^1_0 \, dz z^{n-1} A_{ij}(z,\frac{-p^2}{\mu^2},
\frac{1}{\varepsilon})\,,
\end{eqnarray}
where for simplicity we have not written the moment index $n$ on the
functions.
Also to simplify the expressions we define the phase-space factor
%(A.2)
\begin{eqnarray}
\label{eqn:A.2}
F = \frac{\hat\alpha_s}{4\pi} S_{\varepsilon}
(\frac{-p^2}{\mu^2})^{\varepsilon/2} \,.
\end{eqnarray}

We first split $\hat A_{qq}^{\rm NS}$ into physical
and unphysical parts following the notation in Eq. (\ref{eqn:2.17}).
The physical part is (see also Eq. (\ref{eqn:2.18}))
%(A.3)
\begin{eqnarray}
\label{eqn:A.3}
&&\hat A_{qq}^{\rm NS,PHYS}
\Big(z,\frac{-p^2}{\mu^2},\frac{1}{\varepsilon}\Big)=
\delta(1-z)
\nonumber \\ && \qquad
+F\,\,C_F\Big[\frac{1}{\varepsilon}\Big\{
-4-4z+\frac{8}{1-z}+6\delta(1-z)\Big\}-4+2z
\nonumber \\ && \qquad
+2(\frac{2}{1-z}-1-z)[\ln z+\ln(1-z)]
-\delta(1-z)[7-4\zeta(2)]+\frac{1-\hat{\xi}}{1-z}
\nonumber \\ && \qquad
+\varepsilon\Big\{
1
-(2-z)[\ln z+\ln(1-z)]
+\frac{1}{2}(\frac{2}{1-z}-1-z)[\zeta(2)
\nonumber \\ && \qquad
+\{\ln z+\ln(1-z)\}^2]
+\delta(1-z)[7-\frac{3}{4}\zeta(2)-4\zeta(3)]
+\frac{1}{2}(1-\hat{\xi})
\nonumber \\ && \qquad
\times
[
-\frac{1}{1-z}
+\frac{\ln(1-z)}{1-z}
+\frac{\ln z}{1-z}+\delta(1-z)\{-2+\zeta(2)\}]
\Big\}
\Big]
\nonumber \\ && \qquad
+F^2 \,\, \Big[
\frac{1}{\varepsilon^2}\Big\{
C_F^2[
           40
          + 8z
          - \frac{48}{1-z}
          - 2\delta(1-z)[9-16\zeta(2)]
\nonumber \\ && \qquad
          +32\ln(1-z)(1
          + z
          - \frac{2}{1-z})
       -8 \ln z(
          3
          +3z
          -\frac{4}{1-z}
          )
          ]
\nonumber \\ && \qquad
+C_AC_F[
          \frac{44}{3}
          + \frac{44}{3}z
          - \frac{88}{3}\frac{1}{1-z} 
          - 22\delta(1-z)        
]
\nonumber \\ && \qquad
       +n_fC_FT_f[
          - \frac{16}{3}(1+z-\frac{2}{1-z})
          + 8\delta(1-z)
          ]\Big\}
\nonumber \\ && \qquad
+\frac{1}{\varepsilon}\Big\{
C_F^2[- 56
          +12z+ \frac{56}{1-z}
          +\delta(1-z)[\frac{87}{2}-36\zeta(2)- 8\zeta(3)]
\nonumber \\ && \qquad
          -4\ln(1 - z)(-15+5z+\frac{6}{1-z})
          +24\ln^2(1-z)(1+z-\frac{2}{1-z})
\nonumber \\ && \qquad
          +4\ln z(1+3z-\frac{9}{1-z})
          - 16\ln z\ln(1 - z)\frac{1}{1-z}
\nonumber \\ && \qquad
          -2\ln^2 z(7+7z-\frac{8}{1-z})
          - 8{\rm Li}_2(1 - z)(1+z)]
\nonumber \\ && \qquad
+C_AC_F[\frac{106}{9}
          - \frac{242}{9}z+ \frac{238}{9}\frac{1}{1-z}
          +\delta(1-z)[\frac{325}{6}- \frac{44}{3}\zeta(2)
\nonumber \\ && \qquad
- 12\zeta(3)] 
          +4(1+z-\frac{2}{1-z})[\zeta(2)+\frac{11}{3}\ln(1-z)
-\frac{1}{2}\ln^2z]
\nonumber \\ && \qquad
          + \ln z(\frac{34}{3}+\frac{34}{3}z-
\frac{44}{3}\frac{1}{1-z})
]
\nonumber \\ && \qquad
+n_fC_FT_f[-\frac{8}{9}
          + \frac{40}{9}z
          - \frac{56}{9}\frac{1}{1-z}
          + \delta(1-z)[-\frac{58}{3}+\frac{16}{3}\zeta(2)]
\nonumber \\ && \qquad
          - \frac{8}{3}\{2\ln(1-z)+\ln z\}(1+z
          -\frac{2}{1-z})]\Big\}
\nonumber \\ && \qquad
+C_F^2[\frac{188}{3}-\frac{32}{3}z-\frac{56}{1-z}
          + \zeta(2)(20-4z+ \frac{32}{3}z^2-\frac{4}{1-z})
\nonumber \\ && \qquad
          + 24\zeta(3)(1-z+2z^2-\frac{1}{1-z})
          + \delta(1-z)[-\frac{541}{8}+ \frac{97}{2}\zeta(2)
\nonumber \\ && \qquad
+54\zeta(3)-\frac{74}{5}\zeta(2)^2]
          +\ln(1 - z)(-78+62z+\frac{28}{1-z})
\nonumber \\ && \qquad
          + 8\ln(1 - z)\zeta(2)(1+z-2z^2-\frac{1}{1-z})
          - \ln^2(1-z)(-35+17z
\nonumber \\ && \qquad
+\frac{6}{1-z})
          + \frac{28}{3}\ln^3(1-z)(1+z-\frac{2}{1-z})
\nonumber \\ && \qquad
          - 4\ln(1 - z){\rm Li}_2(1 - z)(3-z+4z^2)
          -\ln z(\frac{44}{3}+\frac{92}{3}z-\frac{44}{1-z})
\nonumber \\ && \qquad
          + \ln z\zeta(2)(2-14z+16z^2)
          + 2\ln z\ln(1 - z)(5+9z-\frac{10}{1-z})
\nonumber \\ && \qquad
          + 2\ln z\ln^2(1-z)(3+3z-\frac{10}{1-z})
          - \ln^2 z(5+7z+\frac{16}{3}z^2
\nonumber \\ && \qquad
-\frac{15}{1-z})
          - \ln^2 z\ln(1 - z)(6-2z+8z^2+\frac{4}{1-z})
          - \ln^3 z(5+5z
\nonumber \\ && \qquad
-\frac{16}{3}\frac{1}{1-z})
          -8 \ln z{\rm Li}_2(1 - z)(1-3z+4z^2+\frac{2}{1-z})
\nonumber \\ && \qquad
          -4 {\rm Li}_2(1 - z)(6-12z+\frac{1}{1-z})
+4{\rm Li}_3(1 - z)(5-7z+12z^2)
\nonumber \\ && \qquad
          + \frac{8}{3}(9+\frac{1}{z}+12z+4z^2)
            [\ln z\ln(1+z)+{\rm Li}_2( - z)]
          -16[\ln z{\rm Li}_2(-z)
\nonumber \\ && \qquad
-2{\rm Li}_3( - z)](1-\frac{1}{1-z})
          -8{\rm S}_{12}(1 - z)(1-7z+6z^2+\frac{4}{1-z})
]
\nonumber \\ && \qquad
+C_AC_F[
          -\frac{1}{27}[745
          -800z+\frac{670}{1-z})
          -\frac{1}{3}\zeta(2)(-23+7z
           +\frac{28}{1-z}
\nonumber \\ && \qquad
+16z^2)
          -2\zeta(3)(10+4z+12z^2-\frac{17}{1-z})
          +\delta(1-z)[-\frac{7081}{72}+\frac{301}{18}\zeta(2)
\nonumber \\ && \qquad
             + 28\zeta(3)+\frac{49}{5}\zeta(2)^2]
          + \frac{2}{9}\ln(1 - z)(71-148z+\frac{119}{1-z})
          + 2\ln(1 - z)
\nonumber \\ && \qquad
\times\zeta(2)(1+3z+4z^2-\frac{5}{1-z})
          + \{\frac{22}{3}\ln^2(1-z)-\ln^3z\}(1+z-\frac{2}{1-z})
\nonumber \\ && \qquad
          -2 \ln z\zeta(2)(3+z+4z^2-\frac{5}{1-z})
 + \frac{1}{6}\ln^2 z(23+71z
            +16z^2-\frac{22}{1-z})
\nonumber \\ && \qquad
          + \frac{1}{9}\ln z(38-64z+\frac{101}{1-z})
          + \frac{2}{3}\ln z\ln(1 - z)(14
            +5z-\frac{19}{1-z})
\nonumber \\ && \qquad
          + 4\ln z{\rm Li}_2(1 - z)(2+4z^2-\frac{3}{1-z})
          -4{\rm Li}_2(1 - z)(3z-\frac{1}{1-z})
\nonumber \\ && \qquad
          -4(3+\frac{1}{3z}+4z+\frac{4}{3}z^2)
             [\ln z\ln(1+z)+{\rm Li}_2( - z)]
\nonumber \\ && \qquad
          +(1-z+4z^2-\frac{1}{1-z})[\ln^2 z\ln(1-z)
          +2\ln(1-z){\rm Li}_2(1-z)
\nonumber \\ && \qquad
-6{\rm Li}_3(1 - z)]
          +8[\ln z{\rm Li}_2(-z)-2{\rm Li}_3( - z)](1-\frac{1}{1-z})
\nonumber \\ && \qquad
          +{\rm S}_{12}(1 - z)(14- 6z+ 24z^2-\frac{14}{1-z})
]
\nonumber \\ && \qquad
+n_fC_FT_f[\frac{4}{27}[
   20-19z +\frac{32}{1-z}]
          + \delta(1-z)[\frac{569}{18}-\frac{46}{9}\zeta(2)
\nonumber \\ && \qquad
-8\zeta(3)]
          +\frac{4}{9}\{2\ln(1-z)+\ln z\}(-1+5z-\frac{7}{1-z})
\nonumber \\ && \qquad
          -\frac{2}{3}[2\zeta(2)+\{2\ln(1-z)+\ln z\}^2](1+z-\frac{2}{1-z})]
\nonumber \\ && \qquad
-2(-1)^n(C_F^2-\frac{1}{2}C_AC_F)\Big(\frac{1}{\varepsilon}\Big\{
     - 8+8z-4\ln z(1+z)
\nonumber \\ && \qquad
     +2(-1+z+\frac{2}{1+z})[2\zeta(2)+4\ln z\ln(1+z)
           -\ln^2 z
           +4{\rm Li}_2(-z)]
\Big\}
\nonumber \\ && \qquad
          \frac{43}{3}(1-z)
          - 8\ln(1 - z)(1-z)
          - 2\zeta(2)(1+7z-\frac{8}{3}z^2)
\nonumber \\ && \qquad
          + 8(z+\frac{1}{1+z})\ln(1+z)[\zeta(2)+\ln z\ln(1+z)
          +2{\rm Li}_2(-z)]
\nonumber \\ && \qquad
          - 4(1+z)[\ln z\ln(1 - z)+{\rm Li}_2(1 - z)+{\rm Li}_3( - z)]
\nonumber \\ && \qquad
          -\frac{1}{3}\ln z(1-11z)
          -\ln^2 z(2-4z+\frac{8}{3}z^2)
          -2\ln^2 z\ln(1 + z)(3- z- \frac{4}{1+z})
\nonumber \\ && \qquad
          - 2\zeta(3)(3+z-\frac{2}{1+z})
     - \frac{4}{3}(12-\frac{1}{z}+9z-4z^2)[\ln z\ln(1+z)+{\rm Li}_2( - z)]
\nonumber \\ && \qquad
          +8{\rm S}_{12}( - z)(-1+ 3z+\frac{4}{1+z})
          +4(1-z-\frac{2}{1+z})[\frac{1}{4}\ln^3 z
\nonumber \\ && \qquad
-\ln z\ln^2(1-z)
           -2\ln z\ln(1 - z)\ln(1 + z)
           +\frac{1}{2}\ln^2 z\ln(1 - z)
\nonumber \\ && \qquad
-2\ln(1 - z){\rm Li}_2(1 - z)
           -2\ln(1 + z){\rm Li}_2(1-z)
           +\ln z{\rm Li}_2(1 - z)
\nonumber \\ && \qquad
-\ln z{\rm Li}_2( - z)+2{\rm Li}_3(1 - z)
           +{\rm S}_{12}(1 - z)+{\rm S}_{12}(z^2)]
\Big)
\Big]\,.
\end{eqnarray}
%(A.4)
Here the factor $(-1)^n$ originates from the non-planar diagrams
(namely 13, 17 and 18 in figure 2 of \cite{mene}). It multiplies
that part of the matrix element which is needed for the mass factorization
of physical processes with two identical quarks in the final state.
The unphysical part (see Eq. (\ref{eqn:2.19})) is equal to
\begin{eqnarray}
\label{eqn:A.4}
&&\hat A_{qq}^{\rm NS,EOM}
\Big(z,\frac{-p^2}{\mu^2},\frac{1}{\varepsilon}\Big)
=
\nonumber \\ && \qquad
F\,\,C_F\Big[4z-2(1-\hat{\xi})
+\varepsilon
[2z-(1-\hat{\xi})][-1 + \ln z+\ln(1-z)]
\Big]
\nonumber \\ && \qquad
+F^2\,\,\Big[\frac{1}{\varepsilon}\Big\{
C_F^2[ -16
          - 8z
          +16z(-2\ln(1 - z)+\ln z)
]
\nonumber \\ && \qquad
+C_AC_F[
\frac{20}{3}
          - \frac{88}{3}z]
+C_FT_f[- \frac{16}{3}
          + \frac{32}{3}z]
\Big\}
\nonumber \\ && \qquad
+C_F^2[
\frac{112}{3}- \frac{88}{3}z
          +\zeta(2)(-8+16z- \frac{32}{3}z^2)
          - 12z\ln(1 - z)
\nonumber \\ && \qquad
          - 24z\ln^2(1-z)
          -\frac{4}{3}\ln z(10-29z)
          + 8(1-4z)\ln z\ln(1 - z)
\nonumber \\ && \qquad
   +\frac{16}{3}(\frac{1}{z}-3z-2z^2)[\ln z\ln(1 + z)+{\rm Li}_2( - z)]
 + \ln^2 z(12z+\frac{16}{3}z^2)
\nonumber \\ && \qquad
          + 16{\rm Li}_2(1 - z)(1-3z)
          +8z(1-z)\{-2\ln(1 - z)\zeta(2)+2\ln z\zeta(2)
\nonumber \\ && \qquad
              -\ln^2 z\ln(1 - z)
-4\ln z{\rm Li}_2(1-z)
              -2\ln(1 - z){\rm Li}_2(1 - z)
\nonumber \\ && \qquad
 +6\zeta(3)+6{\rm Li}_3(1 - z)-6{\rm S}_{12}(1 - z)\}
]
\nonumber \\ && \qquad
+C_AC_F[
          -\frac{86}{9}+\frac{496}{9}z
          +4\zeta(2)(1-2z+\frac{4}{3}z^2)
 -\frac{16}{3}\ln(1-z)(1+4z)
\nonumber \\ && \qquad
  + \ln z(6-28z)
          - \frac{8}{3}z^2\ln^2 z
          - 4[\ln z\ln(1 - z)+2{\rm Li}_2(1 - z)](1-2z)
\nonumber \\ && \qquad
          -\frac{8}{3}(\frac{1}{z}-3z-2z^2)[\ln z\ln(1+z)
            +{\rm Li}_2(-z)]
          + 8z(1-z)\{ -\zeta(2)\ln z
\nonumber \\ && \qquad
\frac{1}{2}\ln^2 z\ln(1-z)
+\zeta(2)\ln(1-z)
+\ln(1-z){\rm Li}_2(1 - z)
\nonumber \\ && \qquad
               +2\ln z{\rm Li}_2(1 - z)
               -3\zeta(3)
-3{\rm Li}_3(1 - z)+3{\rm S}_{12}(1 - z)\}
]
\nonumber \\ && \qquad
+n_fC_FT_f[\frac{8}{9}(5-16z)
          - \frac{8}{3}(1-2z)\{\ln z+2\ln(1-z)\}
]
\nonumber \\ && \qquad
-\frac{16}{3}(-1)^n(C_F^2-\frac{1}{2}C_AC_F)\Big\{
1-z
          +z^2[-2\zeta(2)+\ln^2 z]
\nonumber \\ && \qquad
- \ln z(1-2z)
          +(\frac{1}{z}-3z-2z^2)
           [\ln z\ln(1+z)+{\rm Li}_2( - z)]
\Big\}
\Big]\,.
\end{eqnarray}
%(A.5)
Now we give the pure singlet terms (see Eq. (\ref{eqn:2.21}))
\begin{eqnarray}
\label{eqn:A.5}
&&\hat A_{qq}^{\rm PS, PHYS}
\Big(z,\frac{-p^2}{\mu^2},\frac{1}{\varepsilon}\Big)
=
%\nonumber \\ && \qquad
F^2\,\,n_fC_FT_f\Big[\frac{16}{\varepsilon^2}\Big\{
           5(1-z)
          + 2\ln z(1+z)
          \Big\}
\nonumber \\ && \qquad
       +\frac{8}{\varepsilon}\Big\{
           (1-z)[7
          + 10\ln(1 - z)]
          + \ln z(11+3z)
\nonumber \\ && \qquad
          + 4(1+z)[\ln z\ln(1 - z)+\frac{3}{4}\ln^2 z
              + {\rm Li}_2(1 - z)]
          \Big\}
\nonumber \\ && \qquad
          + 4(1-z)[10
          + 5\zeta(2)
          + 14\ln(1 - z)
          + 10\ln^2(1-z)]
          + 20\ln z(3+z)
\nonumber \\ && \qquad
          + 8\ln z\ln(1 - z)(11+3z)
          + 2\ln^2 z(23+19z)
          + 16{\rm Li}_2(1 - z)(3+4z)
\nonumber \\ && \qquad
          +16(1+z)[2\ln z{\rm Li}_2(1 - z)+\frac{1}{2}\ln z\zeta(2)
           +\frac{3}{2}\ln^2 z\ln(1-z)+\frac{7}{12}\ln^3 z
\nonumber \\ && \qquad
           +\ln z\ln^2(1 - z)
            +2\ln(1-z){\rm Li}_2(1 - z) - 2{\rm Li}_3(1 - z)
\nonumber \\ && \qquad
           +{\rm S}_{12}(1 - z)]
\Big]\,.
\end{eqnarray}
%(A.6)
The result for Eq. (\ref{eqn:2.22}) is
\begin{eqnarray}
\label{eqn:A.6}
&&\hat A_{qq}^{\rm PS, EOM}
\Big(z,\frac{-p^2}{\mu^2},\frac{1}{\varepsilon}\Big)
=
32F^2\,\,n_fC_FT_f\Big[
 \frac{1}{\varepsilon}\Big\{
          -3(1-z)
\nonumber \\ && \qquad
          - \ln z(1+2z)
          \Big\}
 - 3\ln(1 - z)(1-z)
          - \frac{1}{2}\ln z(5+z)
\nonumber \\ && \qquad
          -(1+2z)[\ln z\ln(1 - z)
          +\frac{3}{4}\ln^2 z
          +{\rm Li}_2(1 - z)]
\Big]\,.
\end{eqnarray}
The OME in Eq. (\ref{eqn:2.24}) is equal to
%(A.7)
\begin{eqnarray}
\label{eqn:A.7}
&&\hat A_{qg}^{\rm PHYS}
\Big(z,\frac{-p^2}{\mu^2},\frac{1}{\varepsilon}\Big)
=
F\,\,n_fT_f\Big[-\frac{8}{\varepsilon}[1-2z]
+4-4(1-2z)[\ln z+\ln(1-z)]
\nonumber \\ && \qquad
+\varepsilon\Big\{2\ln(1-z)+2\ln z
          - (1-2z)[\zeta(2)+\{\ln z+\ln(1-z)\}^2]
\Big\}
\Big]
\nonumber \\ && \qquad
+F^2 \,\, n_f
\Big[\frac{8}{\varepsilon^2}\Big\{
C_AT_f[
\frac{2}{3}(25-14z)
          -4\ln(1 - z)  (
          1
          -2z
          )
       + 8\ln z  (
          1
          + z
          )
]
\nonumber \\ && \qquad
+C_FT_f[
3 +2(1-2z)\{-2 \ln(1 - z)+\ln z \} 
]
+n_fT_f^2\frac{8}{3}(1-2z)
\Big\}
\nonumber \\ && \qquad
+\frac{1}{\varepsilon}\Big\{
C_AT_f[
 -\frac{8}{9}(44-z)
       -16\zeta(2)  (
          1
          -4z
          )
  + 64\ln z \ln(1 - z)  (
          1
          +z
          )
\nonumber \\ && \qquad
       -24 \ln^2(1-z)  (
          1
          -2z
          )
       + \frac{8}{3}\ln(1 - z)  (
          73
          - 62z
          )
       + \frac{16}{3}\ln z  (
          14
          +11z
          )
\nonumber \\ && \qquad
 + 8\ln^2 z  (
          5
          +6z
          )
       + 96{\rm Li}_2(1 - z)
       + 16(
          1
          +2z
          )\{\ln z \ln(1+z)+{\rm Li}_2( - z)\}
]
\nonumber \\ && \qquad
+C_FT_f[
 - 4(12-13z)
       + 8\ln(1 - z)  (
          3
          +4z
          )
       -4 \ln z  (
          5
          +8z
          )
       +4(
          1
          -2z
          )
\nonumber \\ && \qquad
\times\{4\zeta(2)-6\ln^2(1-z)-8\ln z \ln(1 - z)
            +3\ln^2 z -12{\rm Li}_2(1 - z)\}
]
\nonumber \\ && \qquad
+n_fT_f^2[-\frac{64}{9}(4-5z)+\frac{32}{3}(1-2z)\{\ln z+\ln(1-z)\}]
\Big\}
\nonumber \\ && \qquad
+C_AT_f[
\frac{4}{27}(1348
          - 1145z)
       - 8\zeta(2)  (
          -6
          + 8z
          + \frac{1}{3}z^2
          )
       -8 \zeta(3)  (
          2
          +z
          -6z^2
          )
\nonumber \\ && \qquad
       -8\ln(1 - z)\zeta(2)  (
          3
          -7z
          +2z^2
          )
       -\frac{28}{3} \ln^3(1-z)  (
          1
          -2z
          )
\nonumber \\ && \qquad
       + \frac{2}{3}\ln^2(1-z)  (
           169
          - 158z
          )
       + 8\ln(1 - z){\rm Li}_2(1 - z)  (
           12
          + 9z
          - 2z^2
          )
\nonumber \\ && \qquad
       -\frac{4}{9} \ln(1 - z)  (
           206
          -145z
          )
       + 8\ln z \zeta(2)  (
          1
          + 3z
          + 2z^2
          )
       + 8\ln z \ln^2(1-z)
\nonumber \\ && \qquad
\times(
          5
          +6z
          )+ \frac{4}{3}\ln z \ln(1 - z)(73+10z)
       + 4\ln^2 z\ln(1 - z)  (
           11
          + 11z
\nonumber \\ && \qquad
       -2z^2)+ \frac{4}{3}\ln^3 z  (
          11
          +14z
          )
       + \frac{4}{3}\ln^2 z  (
          26
          + 47z
          + z^2
          )
\nonumber \\ && \qquad
       + 16\ln z {\rm Li}_2(1 - z)(
          4
          +3z
          -2z^2
          )
       + \frac{16}{9}\ln z  (
           31
          + 58z
          )
\nonumber \\ && \qquad
       -8 {\rm Li}_2(1 - z)  (
          1
          -12z
          )
       -8(
          3
          - \frac{2}{3z}
          + 4z
          + \frac{1}{3}z^2
          )\{\ln z\ln(1+z)
\nonumber \\ && \qquad
+{\rm Li}_2( - z)\}
       -8{\rm Li}_3(1 - z)  (
          8
          +19z
          -6z^2
          )
       -8{\rm S}_{12}(1 - z)  (
           1
          - 13z
          + 6z^2
          )
\nonumber \\ && \qquad
       +8(
          1
          +2z
          )\{2\ln(1+z)\zeta(2)+2\ln(1+z){\rm Li}_2(1-z)
\nonumber \\ && \qquad
            +4\ln(1+z){\rm Li}_2(-z)+2\ln z \ln(1-z)\ln(1+z)
            +2\ln z \ln^2(1+z)
\nonumber \\ && \qquad
+\frac{1}{2}\ln^2 z\ln(1+z)
            +\ln z {\rm Li}_2(-z)-{\rm Li}_3(-z)
            +6{\rm S}_{12}(-z)-{\rm S}_{12}(z^2)\}
]
\nonumber \\ && \qquad
+C_FT_f[
2(23-11z)
       + 2\zeta(2)  (
          3
          - 8z
          )
       + 12\ln^2(1-z)  (
          1
          +2z
          )
\nonumber \\ && \qquad
       -32\ln(1 - z)  (
          2
          -3z
          )
       -4 \ln z \ln(1 - z)  (
          5
          -4z
          )
       -3 \ln^2 z  (
          7
          +8z
          )
\nonumber \\ && \qquad
       +2 \ln z  (
          2
          - 13z
          )
       -32 {\rm Li}_2(1 - z)  (
          1
          -z
          )
       +4(
          1
          -2z
          )\{10\zeta(3)
\nonumber \\ && \qquad
+2\ln(1-z)\zeta(2)-\frac{7}{3}\ln^3(1-z)
-10\ln(1-z){\rm Li}_2(1-z)
\nonumber \\ && \qquad
+5\ln z \zeta(2)-7\ln z\ln^2(1-z)
-5\ln^2 z\ln(1-z)+\frac{7}{6}\ln^3 z
\nonumber \\ && \qquad
-16\ln z {\rm Li}_2(1-z)
+6{\rm Li}_3(1 - z)-12{\rm S}_{12}(1 - z)\}
]
\nonumber \\ && \qquad
+n_fT_f^2\frac{8}{3}[
\frac{86}{9}-\frac{112}{9}z-\frac{8}{3}(4-5z)\{\ln z+\ln(1-z)\}
\nonumber \\ && \qquad
+(1-2z)\{\ln z+\ln(1-z)\}^2
]
\Big]\,.
\end{eqnarray}
The next OME $\hat A_{gq}$ can be split into physical and 
unphysical parts according 
to Eq. (\ref{eqn:2.25}). The former becomes (see Eq. (\ref{eqn:2.26}))
%(A.8)
\begin{eqnarray}
\label{eqn:A.8}
&&\hat A_{gq}^{\rm PHYS}
\Big(z,\frac{-p^2}{\mu^2},\frac{1}{\varepsilon}\Big)
=
F\,\,C_F\Big[
\frac{1}{\varepsilon}
[8-4z]
+(4-2z)[\ln z+\ln(1-z)]+2
\nonumber \\ && \qquad
+\varepsilon\Big\{
(1-\frac{z}{2})[\zeta(2)
+\{\ln(1-z)+\ln z\}^2]
+\ln(1-z)+\ln z\Big\}\Big]
\nonumber \\ && \qquad
+F^2\,\,\Big[\frac{1}{\varepsilon^2}\Big\{
C_F^2[
12z
 + 8(
          2
          -z
          )\{2\ln(1-z)-\ln z\}
]
\nonumber \\ && \qquad
+C_AC_F[
-\frac{8}{3}(14-25z)
  + 16\ln(1 - z)  (
          2
          -z
          )
       -16\ln z  (
          4
          +z
          )
]
\nonumber \\ && \qquad
-n_fC_FT_f\frac{32}{3}(2-z)\Big\}
\nonumber \\ && \qquad
+\frac{1}{\varepsilon}\Big\{
C_F^2[
- 14
          + 4z
       -4\ln(1 - z)  (
          4
          -7z
          )
       - 2z\ln z
      -6(
          2
          -z
          )
\nonumber \\ && \qquad
\times\{-\frac{8}{3}\zeta(2)-2\ln^2(1-z)
+\frac{4}{3}\ln z\ln(1-z)+\ln^2 z+4{\rm Li}_2(1 - z)\}
]
\nonumber \\ && \qquad
+C_AC_F[
  -\frac{4}{9}(47-37z)
       -8\zeta(2)  (
          4
          -z
          )
       +12 \ln^2(1-z)  (
          2
          -z
          )
\nonumber \\ && \qquad
       -\frac{4}{3}\ln(1 - z)  (
          26
          - 49z
          )
       -8 \ln z\ln(1 - z)  (
          2
          +5z
          )
       -4 \ln^2 z  (
          10
          +3z
          )
\nonumber \\ && \qquad
       -\frac{4}{3}\ln z  (
           40
          +7z
          )
       -48z{\rm Li}_2(1 - z)
       -8(
          2
          +z
          )\{\ln z\ln(1+z)+{\rm Li}_2( - z)\}
]
\nonumber \\ && \qquad
-n_fC_FT_f\frac{16}{3}(2-z)[
 \frac{1}{3}
  +\ln(1-z)+2\ln z
]\Big\}
\nonumber \\ && \qquad
+C_F^2[
 25-11z
          -3\zeta(2)(4-3z)
       -16\zeta(3)  (
          1
          -2z
          ) -2\ln(1 - z)  (
          4
          -z
          )
\nonumber \\ && \qquad
       - 6\ln^2(1-z)  (
           2
          -3z
          )
       -8\ln(1 - z){\rm Li}_2(1 - z)  (
          4
          -z
          )
\nonumber \\ && \qquad
       -2\ln z\zeta(2)  (
          2
          -5z
          )
       -4\ln z\ln(1 - z)(1-2z)   + 4\ln(1 - z)\zeta(2)  (
         10
          -7z
          )
\nonumber \\ && \qquad
       -2\ln^2 z\ln(1 - z)  (
          6
          -z
          )
       -\frac{7}{3}(
          2
          -z
          )\{-2\ln^3(1-z)+\frac{12}{7}\ln z\ln^2(1-z)
\nonumber \\ && \qquad
+\ln^3 z- \frac{12}{7}{\rm Li}_2(1 - z)\}
       -\frac{9}{2}z\ln^2 z
       -8 \ln z{\rm Li}_2(1 - z)  (
          2
          +z
          )
\nonumber \\ && \qquad
       - 3\ln z  (
           1
           +2z
          )
       + 24z{\rm Li}_3(1 - z)
       + 12{\rm S}_{12}(1 - z)  (
          2
          -3z
          )
]
\nonumber \\ && \qquad
+C_AC_F[
 -\frac{2}{27}(1400-1573 z)
       -\frac{2}{3}\zeta(2)  (
           2
          - 25z
          + 14z^2
          )
       + 4\zeta(3)  
\nonumber \\ && \qquad
\times (1 -10z          + 6z^2
          )
       -4 \ln(1 - z)\zeta(2)  (
          7
          -7z
          +2z^2
          )
       + \frac{14}{3}\ln^3(1-z)  (
          2
          -z
          )
\nonumber \\ && \qquad
       -\frac{1}{3}\ln^2(1-z)(
           50
          - 97z
          )
       -2(
           7
          + 9z
          + 2z^2
          )\{\ln^2 z\ln(1-z)
\nonumber \\ && \qquad
+2\ln(1 - z){\rm Li}_2(1 - z)\}
       + \frac{8}{9}\ln(1 - z)  (
          13-2z
          )
       -4 \ln z\zeta(2)  (
          5
          +4z
\nonumber \\ && \qquad
     -2z^2)  -2 \ln z\ln^2(1-z)  (
          2
          +15z
          )
       -\frac{4}{3} \ln z\ln(1 - z)  (
          35
          +8z
          )
\nonumber \\ && \qquad
       -\frac{2}{3}\ln^3 z  (
          22
          +7z
          )
       -\frac{1}{3} \ln^2 z  (
           110
          + 71z
          - 14z^2
          )
 -\frac{2}{9} \ln z  (
           421
          + 145z
          )
\nonumber \\ && \qquad
       -8 \ln z{\rm Li}_2(1 - z)  (
          -1
          +2z
          +2z^2
          )
       -4{\rm Li}_2(1 - z)  (
          6
          +11z
          )
\nonumber \\ && \qquad
       -\frac{4}{3}(
          -9
          + \frac{1}{z}
          -3z
          + 7z^2
          )\{\ln z\ln(1+z)+{\rm Li}_2( - z)\}
\nonumber \\ && \qquad
       + 4{\rm Li}_3(1 - z)  (
           13
          - z
          + 6z^2
          )
       + 4{\rm S}_{12}(1 - z)  (
          5
          + 4z
          - 6z^2
          )
\nonumber \\ && \qquad
       + 4(
          2
          +z
          )\{-2 \ln(1 + z)\zeta(2)
       -2 \ln(1 + z){\rm Li}_2(1 - z)
\nonumber \\ && \qquad
       -4\ln(1 + z){\rm Li}_2( - z)
-2\ln z\ln(1-z)\ln(1+z)-2\ln z\ln^2(1+z)
\nonumber \\ && \qquad
-\frac{1}{2}\ln^2 z\ln(1+z)
-\ln z{\rm Li}_2(-z)+{\rm Li}_3( - z)-6{\rm S}_{12}( - z)+{\rm S}_{12}(z^2)\}
]
\nonumber \\ && \qquad
-\frac{4}{3}n_fC_FT_f(2-z)[
\frac{32}{9}+\frac{2}{3}\ln(1-z)+\frac{4}{3}\ln z
+2\zeta(2)
\nonumber \\ && \qquad
+\{\ln(1-z)+2\ln z\}^2
]\Big]\,.
\end{eqnarray}
The unphysical part given by Eq. (\ref{eqn:2.27}) is equal to
%A.9
\begin{eqnarray}
\label{eqn:A.9}
&&\hat A_{gq}^{\rm EOM}
\Big(z,\frac{-p^2}{\mu^2},\frac{1}{\varepsilon}\Big)
=
F\,\,C_F\Big[
-4+4z-2\varepsilon(1-z)[\ln z+\ln(1-z)]\Big]
\nonumber \\ && \qquad
+F^2\,\,\Big[
\frac{16}{\varepsilon}\Big\{
C_F^2[
2-2z+z\ln z
]+n_fC_FT_f\frac{4}{3}(1-z)
\nonumber \\ && \qquad
+C_AC_F[
 \frac{13}{3}(1-z)    -2\ln(1 - z)  (
          1
          -z
          )
       + 2\ln z  (
          2
          +z
          )
]
\Big\}
\nonumber \\ && \qquad
+C_F^2[
- 40(1-z)
       -16(
          1
          - z
          )\{\zeta(2)-\frac{9}{4}\ln(1 - z)\}
       + 16\ln z\ln(1 - z)
\nonumber \\ && \qquad
       + 12z\ln^2 z
       + 16\ln z  (
          1
          -2z
          )
       + 16{\rm Li}_2(1 - z)  (
          2
          -z
          )
]
\nonumber \\ && \qquad
+C_AC_F[
-\frac{148}{9}(1-z)
      +\frac{4}{3}\zeta(2)(
           15
          - 6z
          + 7z^2
          )
       -24 \ln^2(1-z)  (
          1
          -z
          )
\nonumber \\ && \qquad
       + \frac{224}{3}\ln(1 - z)  (
          1
          -z
          )
       + 24\ln z\ln(1 - z)  (
          1
          +3z
          )
 + 16{\rm Li}_2(1 - z)  (
          1
          +5z
          )
\nonumber \\ && \qquad
       + \frac{4}{3}\ln z  (
          33
          +z
          )
       + \frac{4}{3}(
          9
          + \frac{4}{z}
          + 12z
          + 7z^2
          )\{\ln z\ln(1+z)+{\rm Li}_2( - z)\}
\nonumber \\ && \qquad
 + \frac{2}{3}\ln^2 z  (
          63
          +36z
         -7z^2
          )
       + 4(
          1
          -3z
         + 2z^2
          )\{\ln(1-z){\rm Li}_2(1-z)
\nonumber \\ && \qquad
-\ln z\zeta(2)+\ln(1-z)\zeta(2)
+\frac{1}{2}\ln^2 z\ln(1-z)
+2\ln z{\rm Li}_2(1 - z)
\nonumber \\ && \qquad
-3\zeta(3)-3{\rm Li}_3(1 - z)+3{\rm S}_{12}(1 - z)\}
]
\nonumber \\ && \qquad
+n_fC_FT_f\frac{32}{3}(1-z)[
-\frac{2}{3}+\ln(1 - z)+2\ln z
]\Big]\,.
\end{eqnarray}
The gluonic OME in Eq. (\ref{eqn:2.29}) equals
%(A.10)
\begin{eqnarray}
\label{eqn:A.10}
&&\hat A_{gg}^{\rm PHYS}
\Big(z,\frac{-p^2}{\mu^2},\frac{1}{\varepsilon}\Big)
=
\nonumber \\ && \qquad
F\,\,\Big[
\frac{1}{\varepsilon}
\Big\{C_A[8-16z+\frac{8}{1-z}+\frac{22}{3}\delta(1-z)]
-\frac{8}{3}n_fT_f\delta(1-z)\Big\}
\nonumber \\ && \qquad
+C_A[-2+(1-\hat{\xi})\frac{1}{1-z}
+(4-8z+\frac{4}{1-z})\{\ln z+\ln(1-z)\}
\nonumber \\ && \qquad
+\delta(1-z)\{-\frac{67}{9}+4\zeta(2)
-(1-\hat{\xi})+\frac{1}{4}(1-\hat{\xi})^2\}]
+n_fT_f\frac{20}{9}\delta(1-z)
\nonumber \\ && \qquad
+\varepsilon\Big\{C_A[
(1-2z+\frac{1}{1-z})\{\zeta(2)+[\ln z+\ln(1-z)]^2\}
-\ln(1-z)
-\ln z
\nonumber \\ && \qquad
+\frac{1}{2}\frac{1-\hat{\xi}}{1-z}\{-1+\ln z+\ln(1-z)\}
+\delta(1-z)\{\frac{202}{27}-\frac{11}{12}\zeta(2)
-\frac{14}{3}\zeta(3)
\nonumber \\ && \qquad
+\frac{1}{4}\zeta(2)(1-\hat{\xi})
-\frac{1}{4}(1-\hat{\xi})^2\}]
+n_fT_f\delta(1-z)[-\frac{56}{27}+\frac{1}{3}\zeta(2)]
\Big\}\Big]
\nonumber \\ && \qquad
+F^2\,\,\Big[\frac{1}{\varepsilon^2}\Big\{
C_A^2[
 -168+80z
          + \frac{88}{1-z}
 +4\delta(1-z)\{\frac{121}{9}-8\zeta(2)\}
\nonumber \\ && \qquad
       + 64\ln(1 - z)  (
           1
          - 2z
          + \frac{1}{1-z}
          )
       -32 \ln z  (
          3
          +\frac{1}{1-z}
          )
]
\nonumber \\ && \qquad
-n_fC_AT_f 32[
 1-2z+\frac{1}{1-z}
+\frac{11}{9}\delta(1-z)
]
\nonumber \\ && \qquad
+n_fC_FT_f[
80(1-z)
  + 32\ln z  (
          1
          + z
          )
]
+n_f^2 T_f^2\frac{64}{9}\delta(1-z)
\Big\}
\nonumber \\ && \qquad
+\frac{1}{\varepsilon}\Big\{
C_A^2[
\frac{2}{3}(119-29z)
          - \frac{86}{1-z}
       + \delta(1-z)  \{
          - \frac{3326}{27}+ \frac{176}{3}\zeta(2)
\nonumber \\ && \qquad
+20\zeta(3)
          \}
       -8 \zeta(2)  (
          4z
          +\frac{1}{1+z}
          -\frac{1}{1-z}
          )
       + 48\ln^2(1-z)  (
           1
          - 2z
\nonumber \\ && \qquad
          + \frac{1}{1-z}
          )
       + \frac{16}{3}\ln(1 - z)  (
          - 40
          + 26z
          + \frac{11}{1-z}
          )
\nonumber \\ && \qquad
       -16 \ln z\ln(1 - z)  (
           3
          + 6z
          - \frac{1}{1-z}
          )
       -4 \ln^2 z  (
           16
          - \frac{1}{1+z}
          + \frac{5}{1-z}
          )
\nonumber \\ && \qquad
       - \frac{4}{3}\ln z  (
           57
          + 21z
          - \frac{44}{1-z}
          )
       -64 {\rm Li}_2(1 - z)  (
          1
          +z
          )
\nonumber \\ && \qquad
       -16(
           1
          + 2z
          + \frac{1}{1+z}
          )\{\ln z\ln(1 + z)+ {\rm Li}_2( - z)\}
]
\nonumber \\ && \qquad
+n_fC_AT_f[
\frac{16}{3}(10-13z)
+ \frac{24}{1-z}
       + \frac{8}{3}\delta(1-z)  \{
          \frac{271}{9}-8\zeta(2)
          \}
\nonumber \\ && \qquad
       -\frac{64}{3} \ln(1 - z)  (
          1
          -2z
          +\frac{1}{1-z}
          )
       -16\ln z  (
          1
          -3z
          +\frac{4}{3}\frac{1}{1-z}
          )
]
\nonumber \\ && \qquad
+n_fC_FT_f[
 - 88(1-z)+4\delta(1-z)
 + 80\ln(1 - z)  (
           1
          - z
          )
\nonumber \\ && \qquad
       + 8\ln z  (
          3
          - 7z
          )
  + 8(
          1
          + z
          )\{4\ln z\ln(1 - z)+3\ln^2 z+4{\rm Li}_2(1 - z)\}
]
\nonumber \\ && \qquad
-n_f^2 T_f^2\frac{320}{27}\delta(1-z)
\Big\}
\nonumber \\ && \qquad
+C_A^2[
 -\frac{4}{9}(565-403z)
          + \frac{254}{3}\frac{1}{1-z}
       - \zeta(2)  (
           \frac{149}{3}
          - \frac{181}{3}z
          + 8z^2
\nonumber \\ && \qquad
          - \frac{19}{1-z}
          )
       + \delta(1-z)  \{
          \frac{11141}{54} - \frac{214}{3}\zeta(2)
            -\frac{854}{9}\zeta(3)+ 5\zeta(2)^2
          \}
\nonumber \\ && \qquad
       + \zeta(3)  (
          - 14
          + 52z
          + 15z^2
          - \frac{4}{1+z}
          - \frac{10}{1-z})
\nonumber \\ && \qquad
       + \ln(1 - z)\zeta(2)  (
           10
          - 28z
          - 5z^2
          + \frac{18}{1-z}
          )
       + \frac{56}{3}\ln^3(1-z)
\nonumber \\ && \qquad
\times  (
          1
          - 2z
          + \frac{1}{1-z}
          )
       +2\ln^2(1-z)  (
          - 59
          + 42z
          + \frac{11}{1-z}
          )
\nonumber \\ && \qquad
       - \ln(1 - z){\rm Li}_2(1 - z)  (
          74
          +108z
          + 5z^2
          + \frac{16}{1+z}
          - \frac{2}{1-z}
          )
\nonumber \\ && \qquad
       -\frac{1}{9} \ln(1 - z)  (
          - 1267
          + 959z
          + \frac{506}{1-z}
          )
\nonumber \\ && \qquad
       -4 (
           5
          + 8z
          + \frac{2}{1+z}
         )\{\ln(1+z)\zeta(2)+2\ln(1 + z){\rm Li}_2( - z)\}
\nonumber \\ && \qquad
       - \ln z\zeta(2)  (
           30
          - 12z
          - 5z^2
          + \frac{10}{1-z}
          )
       -4 \ln z\ln^2(1-z)  (
           5
          + 22z
\nonumber \\ && \qquad
          + \frac{2}{1+z}
          - \frac{5}{1-z}
          )
       +\frac{1}{3} \ln z\ln(1 - z)  (
          - 245
          -29z
          + \frac{97}{1-z}
          )
\nonumber \\ && \qquad
       -4 \ln z\ln^2(1+z)  (
           5
          +8z
          + \frac{2}{1+z}
          )
       -2 \ln^2 z\ln(1 + z)  (
           1
          + 4z
\nonumber \\ && \qquad
          + \frac{4}{1+z}
          )
       - \ln^2 z\ln(1 - z)  (
           37
          + 54z
          + \frac{5}{2}z^2
          - \frac{4}{1+z}
          - \frac{5}{1-z}
          )
\nonumber \\ && \qquad
       -2 \ln^3 z  (
           12
          -\frac{1}{1+z}
          + \frac{11}{3}\frac{1}{1-z}
          )
       -\ln^2 z  (
           31
          + 41z
          - 4z^2
          - \frac{22}{1-z}
          )
\nonumber \\ && \qquad
       - 2\ln z{\rm Li}_2(1 - z)  (
           10
          + 60z
          + 5z^2
          - \frac{4}{1+z}
          - \frac{14}{1-z}
          )
\nonumber \\ && \qquad
       -4 \ln z{\rm Li}_2( - z)  (
           1
          + 4z
          + \frac{2}{1+z}
          + \frac{2}{1-z}
          )
       -\ln z  (
           \frac{226}{3}
          + 74z
\nonumber \\ && \qquad
          + \frac{521}{9}\frac{1}{1-z}
          )
       -2 {\rm Li}_2(1 - z)  (
          - 21
          + 45z
          + \frac{13}{3}\frac{1}{1-z}
          )
\nonumber \\ && \qquad
       + 4(
           8
          - \frac{1}{z}
          + 7z
          - 2z^2
          )\{\ln z\ln(1+z)+{\rm Li}_2( - z)\}
\nonumber \\ && \qquad
       + {\rm Li}_3(1 - z)  (
           62
          + 132z
          + 15z^2
          + \frac{16}{1+z}
          - \frac{6}{1-z}
          )
\nonumber \\ && \qquad
       + 4{\rm Li}_3( - z)  (
          1
          + 4z
          + \frac{4}{1-z}
          )
       + {\rm S}_{12}(1 - z)  (
           46
          - 148z
          - 15z^2
\nonumber \\ && \qquad
          + \frac{8}{1+z}
          + \frac{46}{1-z}
          )
       -8 {\rm S}_{12}( - z)  (
           7
          + 12z
          + \frac{4}{1+z}
          )
       + 8(
           1
          + 2z
          + \frac{1}{1+z}
          )
\nonumber \\ && \qquad
\times\{-2\ln(1+z){\rm Li}_2(1-z)
-2\ln z\ln(1-z)\ln(1+z)
                +{\rm S}_{12}(z^2)\}
]
\nonumber \\ && \qquad
+n_fC_AT_f[
-\frac{2}{9}(473
          - 563z)
         -\frac{64}{3}\frac{1}{1-z}
       + \frac{4}{3}\zeta(2)  (
           13
          + 10z
          + 4z^2
\nonumber \\ && \qquad
          - \frac{6}{1-z}
          )
       + \delta(1-z)  \{
     - \frac{3224}{27}+\frac{236}{9}\zeta(2)+ \frac{136}{9}\zeta(3)
          \}
\nonumber \\ && \qquad
       - 8\ln^2(1-z)  (
           1
          - 2z
          +\frac{1}{1-z}
          )
       + \frac{4}{9}\ln(1 - z)  (
           97
          - 131z
\nonumber \\ && \qquad
          + \frac{34}{1-z}
          )
       -\frac{4}{3} \ln z\ln(1 - z)  (
           11
          - 22z
          + \frac{8}{1-z}
          )
\nonumber \\ && \qquad
       -4\ln^2 z  (
           3
          - 5z
          + \frac{2}{3}z^2
          + \frac{2}{1-z}
          )
       + \frac{4}{3}\ln z  (
           10
          - 37z
          + \frac{37}{3}\frac{1}{1-z}
          )
\nonumber \\ && \qquad
       -\frac{8}{3}{\rm Li}_2(1 - z)  (
          3
          - \frac{2}{1-z}
          )
       + \frac{16}{3}(
          3
          + \frac{1}{z}
          + 3z
          + z^2
          )\{\ln z\ln(1+z)
\nonumber \\ && \qquad
+{\rm Li}_2( - z)\}
       +8z^2\{\ln(1 - z){\rm Li}_2(1 - z)
            -\ln z\zeta(2)
+\frac{1}{2}\ln^2 z\ln(1 - z)
\nonumber \\ && \qquad
+\ln(1- z)\zeta(2)
            +2\ln z{\rm Li}_2(1 - z)
             -3\zeta(3)-3{\rm Li}_3(1 - z) + 3{\rm S}_{12}(1 - z)\}
]
\nonumber \\ && \qquad
+n_fC_FT_f[\delta(1-z)\{ -\frac{55}{3} + 16 \zeta(3)\} + 
\frac{568}{3}(1-z)
       - 4\zeta(2)  (
           3
          + 5z
          + \frac{8}{3}z^2
          )
\nonumber \\ && \qquad
       + 40\ln^2(1-z)  (
           1
          - z
          )
       -88 \ln(1 - z)  (
          1
          -z
          )
       + 8\ln z\ln(1 - z)  (
          3
          - 7z
          )
\nonumber \\ && \qquad
       + 2\ln^2 z  (
           7
          - 11z
          + \frac{8}{3}z^2
          )
       -\frac{4}{3} \ln z  (
           1
          - 65z
          )
       -\frac{32}{3}(
          3
          +\frac{1}{z}
          +3z
          +z^2
          )
\nonumber \\ && \qquad
\times \{\ln z\ln(1+z)+{\rm Li}_2( - z)\}
       + 16(
          1
          +z
          )(2\ln(1-z){\rm Li}_2(1-z)
\nonumber \\ && \qquad
            +\frac{1}{2}\ln z\zeta(2)+\ln z\ln^2(1-z)
            +\frac{3}{2}\ln^2 z\ln(1 - z)
            +\frac{7}{12}\ln^3 z
\nonumber \\ && \qquad
+2\ln z{\rm Li}_2(1 - z)-{\rm Li}_2(1 - z)
            -2{\rm Li}_3(1 - z)+{\rm S}_{12}(1 - z))
]
\nonumber \\ && \qquad
+n_f^2 T_f^2 16\delta(1-z)[1-\frac{1}{9}\zeta(2)]
\Big]\,.
\end{eqnarray}
Finally we have determined the  unrenormalized $\hat Z_{qq}^{5,{\rm NS}}$ from
Eq. (\ref{eqn:2.15}). The numerator, which is equal to the spin averaged
OME $\hat A_{qq}^{\rm NS,PHYS}$, is given in Eq. (A.3) of \cite{msn}. The 
denominator has been computed above (see Eq. (\ref{eqn:A.3})).
The result is equal to
%(A.11)
\begin{eqnarray}
\label{eqn:A.11}
\hat Z_{qq}^{5,{\rm NS}}&=& \delta(1-z) + \hat a_s S_\varepsilon 
C_F \Big [ - 8 (1-z) +
\varepsilon \Big \{ - 4(1-z)(\ln(1-z)+\ln z)
\nonumber\\[2ex]
&& + 2z - (1-\hat{\xi}) \Big \} \Big ]
\nonumber\\[2ex]
&& + \hat a_s^2 S_\varepsilon^2 \Big [ C_F^2 \Big \{
 - 16 (1 - z) - 8(2 + z)\ln z + 16(1-z)\ln z \ln(1-z) \Big \}
\nonumber\\[2ex]
&& + C_A C_F \Big \{  \frac{1}{\varepsilon}\frac{88}{3}(1-z)
-\frac{562}{9} + \frac{460}{9}z + 8(1-z)\zeta(2)
\nonumber\\[2ex]
&&+ \frac{88}{3}(1-z)\ln(1-z)
+(\frac{8}{3}-\frac{80}{3}z)\ln z -4(1-z)\ln^2 z \Big \}
\nonumber\\[2ex]
&& + n_f C_F T_f \Big \{
- \frac{1}{\varepsilon}\frac{32}{3}(1-z) + \frac{56}{9}
- \frac{32}{9} z - \frac{32}{3}(1-z)\ln(1-z)
\nonumber\\[2ex]
&& - \frac{16}{3}(1-z)\ln z \Big \}
\nonumber\\[2ex]
&& + (-1)^n (C_F^2-\frac{1}{2}C_A C_F) \Big \{  8(1+z) \Big ( 4 {\rm Li}_2(-z)
+ 4\ln z \ln (1+z)
\nonumber\\[2ex]
&& + 2\zeta(2) - \ln^2 z - 3\ln z \Big ) - 56(1-z) \Big \} \Big ]\,.
\end{eqnarray}
%%%%%%%%%%%%%%%%%%%%% REFERENCES %%%%%%%%%%%%%%%%%%%%%%%%%%%%%%%%%%%%%%%%%%%%%
%\begin{thebibliography}{99}

\end{document}